\pdfoutput=1

\documentclass[10pt,aps,prx,twocolumn,superscriptaddress,notitlepage,longbibliography,floatfix]{revtex4-2}

\usepackage{amsmath}
\usepackage{amsfonts}
\usepackage{amssymb}
\usepackage{xcolor}
\usepackage[T1]{fontenc}
\definecolor{Blue}{HTML}{2D2F92}
\definecolor{Gray}{HTML}{AAAAAA}
\usepackage{booktabs}

\usepackage{newtxtext}
\usepackage{newtxmath}
\usepackage{comment}

\usepackage{tabularx}
\usepackage{mathtools}
\usepackage{graphicx}
\usepackage{bm}
\usepackage{placeins}
\usepackage{siunitx}

\usepackage[toc,page,header]{appendix}

\allowdisplaybreaks

\usepackage[colorlinks=true,allcolors=Blue]{hyperref}
\usepackage[normalem]{ulem}

\begin{document}
\hbadness=100000
\hfuzz=10000pt

\title{Renormalized mechanics and stochastic thermodynamics of growing vesicles}

\author{Jordan L.\ Shivers}
\affiliation{The James Franck Institute, University of Chicago, Chicago, Illinois USA}
\affiliation{Department of Chemistry, University of Chicago, Chicago, Illinois USA}
\author{Michael Nguyen}
\affiliation{The James Franck Institute, University of Chicago, Chicago, Illinois USA}
\affiliation{Department of Chemistry, University of Chicago, Chicago, Illinois USA}
\author{Aaron R. Dinner}
\affiliation{The James Franck Institute, University of Chicago, Chicago, Illinois USA}
\affiliation{Department of Chemistry, University of Chicago, Chicago, Illinois USA}
\author{Petia M. Vlahovska}
\affiliation{Engineering Sciences and Applied Mathematics, Northwestern University, Evanston, Illinois USA}
\author{Suriyanarayanan Vaikuntanathan}
\email{svaikunt@uchicago.edu}
\affiliation{The James Franck Institute, University of Chicago, Chicago, Illinois USA}
\affiliation{Department of Chemistry, University of Chicago, Chicago, Illinois USA}

\begin{abstract}

Uncovering the rules governing the nonequilibrium dynamics of the membranes that define biological cells is of central importance to understanding the physics of living systems. We theoretically and computationally investigate the behavior of flexible quasispherical vesicles that exchange membrane constituents, internal volume, and heat with an external reservoir. The excess chemical potential and osmotic pressure difference imposed by the reservoir act as generalized thermodynamic driving forces that modulate vesicle morphology. We show that the renormalization of membrane mechanical properties by nonequilibrium driving gives rise to a morphological transition between a weakly driven regime, in which growing vesicles remain quasispherical, and a strongly driven regime, in which vesicles accommodate rapid membrane uptake by developing surface wrinkles. Additionally, we propose a minimal vesicle growth-shape law, derived using insights from stochastic thermodynamics, that robustly describes vesicle growth dynamics even in strongly driven, far-from-equilibrium regimes.

\end{abstract}

\maketitle

\vspace{1em}

\section{Introduction}

\begin{figure}[htbp!]
   \centering
   \includegraphics[width=3.375in]{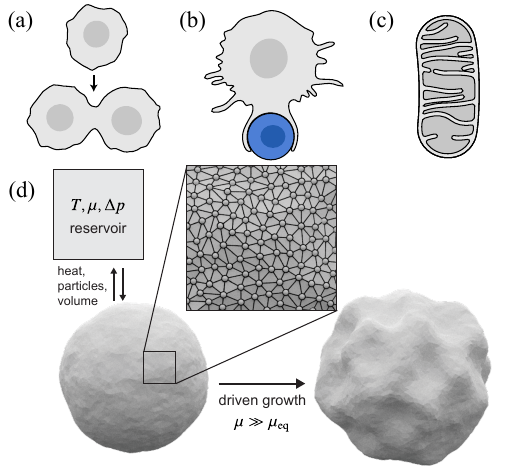}
   \caption{\textbf{Examples of membrane growth processes in living cells and schematic representation of our Monte Carlo vesicle model.} Membrane growth is central to the dynamic structure and function of cells and their internal compartments, and it plays a key role in active processes such as (a) cell division, (b) phagocytosis of pathogens by macrophages, and (c) the shape dynamics of organelles such as mitochondria. (d) Schematic of our Monte Carlo simulations for $d=3$. A fluctuating vesicle is modeled as a quasispherical triangulated mesh that exchanges heat, surface particles (vertices), and volume with a reservoir characterized by temperature $T$, chemical potential $\mu$, and osmotic pressure difference $\Delta p$. For strongly nonequilibrium growth conditions ($\mu \gg \mu_\text{eq}$), we observe a morphological transition between a near-equilibrium regime in which the shapes of growing vesicles remain quasispherical and a far-from-equilibrium regime with persistent wrinkling.}
   \label{fig:fig_1}
\end{figure}

Membranes are essential components of living cells, functioning as flexible physical and chemical barriers that compartmentalize cellular contents \cite{fawcett_cell_1981,alberts_molecular_2022}. Beyond this basic role, membranes perform a multitude of additional functions, enabled by their remarkable physical properties: they flow like two-dimensional fluids while also exhibiting out-of-plane bending elasticity.
\cite{helfrich_elastic_1973, deuling_curvature_1976,phillips_physical_2013}. When combined with nonequilibrium processes at the membrane like pump activity and  active fission/fusion \cite{rao_active_2001, ramaswamy_physics_2001, rautu_active_2024}, these properties enable diverse membrane morphologies and dynamics that are critical for cellular function (Fig.\ \ref{fig:fig_1}a-c). The ability of membranes to grow and change shape may, in fact, have played a pivotal role in life's origins \cite{szostak_synthesizing_2001,hanczyc_experimental_2003,chen_membrane_2004,zhu_coupled_2009}. Simple fatty acid vesicles capable of growth and division through membrane assembly dynamics could have provided a primitive form of cellular compartmentalization on the early Earth \cite{schrum_origins_2010,budin_concentration-driven_2012}. While the growth and division of modern cells involves complex regulatory mechanisms that coordinate changes in surface area and volume, early protocells lacked these mechanisms and thus must have relied on intrinsic physical and chemical processes to proliferate \cite{imai_giant_2019,imai_vesicles_2022,lipowsky_remodeling_2022}. 

In modern cells, active processes regulate the size and shape of the plasma membrane and internal membrane-bound organelles \cite{lippincott-schwartz_secretory_2000,luzio_membrane_2003,vedrenne_morphogenesis_2006,saftig_lysosome_2009,lamb_autophagosome_2013,sens_remodeling_2013,luzio_biogenesis_2014,carlsson_membrane_2015,stefan_membrane_2017,giacomello_cell_2020,nakatogawa_mechanisms_2020,tabara_molecular_2025}. Active processes can strongly influence membrane mechanics \cite{prost_shape_1996, prost_active_1998, turlier_unveiling_2019}, leading to dramatic changes in properties such as the membrane tension \cite{manneville_activity_1999, girard_passive_2005,solon_negative_2006,faris_membrane_2009,sinha_cells_2011,loubet_effective_2012, sens_membrane_2015,lavi_cellular_2019}. In particular, the driven uptake of surface material from an external reservoir into an otherwise passive membrane leads to a reduced or even negative effective tension \cite{girard_fluid_2004}; in experiments on initially quasispherical vesicles, this produces large-amplitude shape fluctuations that reflect localized mechanical instabilities \cite{solon_negative_2006}. 
Membrane tension is a regulator of many cellular processes, such as growth \cite{morris_cell_2001,mccusker_plasma_2012}, division \cite{lecuit_polarized_2000,albertson_membrane_2005,mccusker_cdk1_2007,boucrot_endosomal_2007,carlton_membrane_2020,frey_membrane_2022}, motility \cite{roubinet_common_2012,tanaka_turnover_2017,hetmanski_membrane_2019,de_belly_cell_2023,ghisleni_mechanotransduction_2024}, wound repair \cite{raj_membrane_2024}, endo- and exocytosis \cite{gauthier_plasma_2009,heinrich_blurred_2011,sudhof_synaptic_2011,diz-munoz_use_2013,walani_endocytic_2015}, organelle dynamics \cite{upadhyaya_tension_2004,mierke_mechanical_2020,lipowsky_elucidating_2023}, and adaptation to osmotic stress \cite{dai_membrane_1998,pullarkat_osmotically_2006}. Membrane curvature is also modulated by activity \cite{bassereau_2018_2018,turlier_unveiling_2019}.
Yet, the dependence of tension and bending rigidity on nonequilibrium, active processes remains elusive. 

Here, we investigate membrane growth dynamics and associated changes in membrane elastic properties using nonequilibrium Monte Carlo simulations of growing vesicles that exchange membrane material and volume with their environment (Fig.\ \ref{fig:fig_1}d). In our simulations, imposed gradients of chemical potential and osmotic pressure drive fluxes of membrane material (surface particles) and internal volume from external reservoirs into a fluctuating vesicle, giving rise to morphological changes that reflect the interplay of driven growth and energetically costly deformations. We generate ensembles of growth trajectories over a range of nonequilibrium conditions and demonstrate that the configurational distributions are amenable to an effective equilibrium mechanical description, in which the effective mechanical properties are renormalized by nonequilibrium driving. Notably, we observe that, within this description, increasing the excess chemical potential can cause the effective tension to cross a theoretically predicted critical threshold---derived for equilibrium elastic shells under external pressure \cite{kosmrlj_statistical_2017}---beyond which a buckling instability is expected.  The onset of buckling predicted by this theory is consistent with the transition from stable to unstable growth---and the onset of wrinkling---that we observe in our simulations.

We then examine the system through the lens of stochastic thermodynamics \cite{seifert_stochastic_2012,seifert_stochastic_2019,peliti_stochastic_2021}, measuring distributions of observed fluxes---e.g., of particles and volume from the reservoir to the vesicle----over ensembles of trajectories. We interpret these in the context of existing hydrodynamic theories built upon linear irreversible thermodynamics in Refs.\ \onlinecite{girard_fluid_2004,salbreux_mechanics_2017,sahu_irreversible_2017,julicher_hydrodynamic_2018}. Guided by this, we show how this information can inform simple \emph{low-dimensional growth laws} that describe relationships between the net fluxes (of area or volume), their fluctuations, and the associated driving forces.  These results highlight the interplay between thermodynamic driving and large-scale morphological changes, providing a starting point for describing more complex, actively growing membrane systems across scales, from organelles to protocells.

\section{Monte Carlo simulations} \label{sec:monte_carlo}

We develop nonequilibrium Monte Carlo simulations of the  growth of closed membranes (vesicles) in two and three dimensions ($d=2,3$). Here we outline the approach, while  the technical details are provided in Appendix \ref{app:simulation_details}. 

We model fluid vesicles in $d=3$ as quasispherical triangulated surfaces \cite{gompper_network_1997,nelson_statistical_2004}, as depicted in Fig.\ \ref{fig:fig_1}d. Triangulated mesh-based models have been used extensively for the mesoscopic modeling of membranes in both equilibrium and nonequilibrium contexts \cite{ramakrishnan_membrane-mediated_2013,gueguen_fluctuation_2017,drab_monte_2021}, including simulations of the growth of fluid membranes driven by active recycling \cite{sachin_krishnan_active_2022}, tension gradients \cite{okuda_continuum_2022}, and filament polymerization \cite{weichsel_more_2016}, the growth of elastic shells \cite{rotskoff_robust_2018, li_chemically_2021}, and the dynamics of both inactive and active elastic shells under pressure \cite{paulose_fluctuating_2012,agrawal_active_2023}. In these inherently coarse-grained models, each vertex (or each face) represents  a patch of a membrane bilayer composed of many constituent molecules \cite{li_mesoscopic_2024, dasanna_mesoscopic_2024}.  Within this mesh representation, the configuration of a given vesicle is specified by a set of $N$ vertex positions $\left\{\bm{r}\right\}$ and the mesh adjacency matrix $\bm{A}$, in which $A_{ij} = 1$ if vertex $i$ is connected to vertex $j$ and $0$ otherwise. In our simulations, the total energy for a configuration, $E_\mathrm{tot}$, is given by
\begin{equation} \label{eq:energy}
E_\mathrm{tot} = E_\mathrm{bend} + E_\mathrm{tether} + E_\mathrm{area},
\end{equation}
in which $E_\mathrm{bend}$ is the bending energy, $E_\mathrm{tether}$ constrains the distance between vertices connected by an edge, and $E_\mathrm{area}$ is the local area constraint energy \cite{vutukuri_active_2020}. These are defined in Appendix \ref{app:simulation_details_3D}. 

We simulate a fluid vesicle in contact with an external reservoir of surface particles with chemical potential $\mu$, a heat bath with temperature $T$, and a volume reservoir characterized by an osmotic pressure difference $\Delta p=p_\mathrm{in}-p_\mathrm{out}$. The number of surface particles (vertices), $N$, and the vesicle volume, $V$, fluctuate in a manner governed by the specified chemical potential and osmotic pressure difference, respectively.  We denote the instantaneous state of the system by $\Gamma = \left[\left\{\bm{r}\right\}, N, V\right]$, with which we associate a function $\Phi(\Gamma)\equiv\Phi(\Gamma,\mu,\Delta p)$ defined as
\begin{equation}
    \Phi(\Gamma) = E_\mathrm{tot}(\left\{\bm{r}\right\}) - \mu N - \Delta p V.
\end{equation}

We evolve the system stochastically using the Monte Carlo method, with a move set that includes vertex translation, edge flipping, vertex addition ($N\to N+1$), and vertex removal ($N \to N-1$) moves. 

The probability of a transition from state $\Gamma$ to state $\Gamma'$ is given by $W(\Gamma\to\Gamma')$, defined as
\begin{equation}
    W(\Gamma\to\Gamma') = P_\mathrm{gen}(\Gamma\to\Gamma') P_\mathrm{acc}(\Gamma\to\Gamma')
\end{equation}
in which $P_\mathrm{gen}(\Gamma\to\Gamma')$ and $P_\mathrm{acc}(\Gamma\to\Gamma')$ are the probabilities of proposing and accepting, respectively, the transition $\Gamma\to\Gamma'$.
The ratio of the forward and reverse transition probabilities depends on the associated change in $\Phi(\Gamma)$ as
\begin{equation}
    \frac{W(\Gamma\to\Gamma')}{W(\Gamma'\to\Gamma)} = \exp\left(-\frac{\Delta\Phi(\Gamma\to\Gamma')}{k_\mathrm{B} T}\right),
\end{equation}
in which $\Delta \Phi(\Gamma \to \Gamma') = \Phi(\Gamma') - \Phi(\Gamma)$, $k_\mathrm{B}$ is Boltzmann's constant, and $T$ is the temperature.
. To this end, proposed state transitions $\Gamma\to\Gamma'$ are accepted according to the Metropolis criterion \cite{metropolis_equation_1953} with probability $P_\mathrm{acc}(\Gamma\to\Gamma')$ given by
\begin{equation} \label{eq:metropolis}
    P_\mathrm{acc}(\Gamma \to \Gamma') = \min\left[1, \frac{P_\mathrm{gen}(\Gamma'\to\Gamma)}{P_\mathrm{gen}(\Gamma\to\Gamma') }\exp\left(-\frac{\Delta \Phi(\Gamma\to\Gamma')}{k_\mathrm{B} T}\right)\right].
\end{equation}
For a given value of the osmotic pressure $\Delta p$ and temperature $T$, there is a particular \textit{equilibrium} chemical potential $\mu=\mu_\mathrm{eq}$ at which this procedure samples an equilibrium configurational probability distribution $p_\mathrm{eq}(\Gamma)$ given by
\begin{equation}
    p_\mathrm{eq}(\Gamma) = \frac{1}{Z} \exp\left[-\frac{1}{k_\mathrm{B} T}\Phi_\mathrm{eq}(\bm{r}, N, V)\right],
\end{equation}
in which $\Phi_\mathrm{eq}(\Gamma) = \Phi(\Gamma,\mu_\mathrm{eq},\Delta p)$, and $Z$ is the corresponding partition function. The inclusion of the proposal probabilities in Eq.~\ref{eq:metropolis} ensures that, for $\mu=\mu_\mathrm{eq}$, detailed balance is preserved, i.e.,
\begin{equation}
     p_\mathrm{eq}(\Gamma) W(\Gamma\to\Gamma') = p_\mathrm{eq}(\Gamma') W(\Gamma'\to\Gamma) .
\end{equation}
For $\mu \neq \mu_\mathrm{eq}$, detailed balance is broken, and our Monte Carlo procedure no longer samples a stationary probability distribution. That is, for $\mu > \mu_\mathrm{eq}$, transitions that increase the number of vertices, $N\to N+1$, become more favorable than the reverse; the opposite is true for $\mu < \mu_\mathrm{eq}$. 

Simulations are initialized with an approximately spherical vesicle of average radius $R_0=15\ell_0$, with edges of average length $\ell_0 \equiv \langle \ell_{ij} \rangle \approx 1$. For all of our simulations, we set the vesicle bending rigidity to $\kappa = 20 k_\mathrm{B} T$, within the range of typical values for biological membranes \cite{phillips_emerging_2009,phillips_membranes_2018}. In each Monte Carlo sweep, we attempt exactly $N$ vertex translation moves, an average of $n_\mathrm{edges} p_\mathrm{flip}$ edge flips, and an average of $N p_\mathrm{exchange}$ vertex addition and removal moves, each. Here, $n_\mathrm{edges}$ is the number of edges in the mesh, $p_\mathrm{flip}$ is the probability of attempting an edge flip for each edge per sweep, and $p_\mathrm{exchange}$ is the particle exchange attempt rate.  We first equilibrate the system at the equilibrium chemical potential $\mu_\mathrm{eq}$, determined a priori, for $n_\mathrm{equil}$ sweeps. After equilibration is complete, we change the reservoir chemical potential to the target value, $\mu$, and perform $n_\mathrm{sim}$ sweeps. Further implementation details are given in Appendix \ref{app:monte_carlo_3D}.

The setup for simulations in $d=2$ is described in Appendix \ref{app:simulation_details_2D}.

\begin{figure*}[htbp!]
   \centering
   \includegraphics[width=7in]{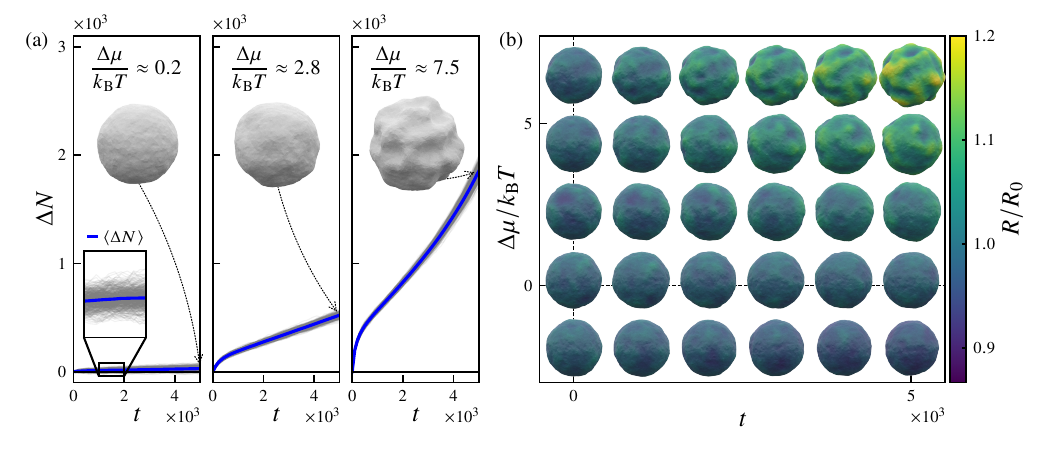}
   \caption{\textbf{Phenomenology of vesicle growth dynamics for various imposed excess chemical potentials.} (a) Ensembles of growth trajectories for various values of excess chemical potential $\Delta\mu = \mu - \mu_\mathrm{eq}$, with the net influx of surface particles (vertices) ($\Delta N = N(t) - N(0)$) plotted as a function of the number of elapsed Monte Carlo sweeps $t$. Light gray lines correspond to individual trajectories and blue lines correspond to the ensemble average $\langle \Delta N(t) \rangle$. (b) Representative snapshots of vesicle configurations after number of sweeps $t$, for different values of the excess chemical potential $\Delta\mu/k_\mathrm{B}T$. Images are centered on the corresponding values of $(t,\;\Delta\mu)$. The vesicle surface is colored by the normalized radius $R/R_0$, in which $R_0$ is the average radius at $t=0$. 
   For sufficiently large nonequilibrium driving (large $\Delta\mu$), growing vesicles exhibit highly deformed morphologies, with significant out-of-plane undulations.  For these simulations, the particle reservoir exchange attempt rate is $p_\mathrm{exchange} = 1$, the imposed osmotic pressure difference is $\Delta p = 0$, and the number of samples is $n_\mathrm{samples}=500$.}
   \label{fig:fig_2}
\end{figure*}

\section{Results}

\subsection{Vesicle growth dynamics and morphology} \label{sec:growth_dynamics}

We begin by mapping the dependence of the dynamics of vesicle growth and the associated morphological changes on the imposed chemical potential $\mu$ and the osmotic pressure $\Delta p$. For a given set of conditions, we generate an ensemble of $n_\mathrm{samples}=500$ growth trajectories. For each trajectory, we compute the associated net fluxes given by $\bm{J}(t) = \Delta \bm{x}(t) \equiv \bm{x}(t) - \bm{x}(0)$ in which $\bm{x} = (N,\;V,\;...)^T$ denotes the set of observables and $t$ refers to the number of elapsed Monte Carlo sweeps.  We refer to the variable $t$ as time, measured in Monte Carlo sweeps; however, one should note that this is not a true measure of time, and the ``dynamics'' that we observe are not guaranteed to be physically accurate. Nonetheless, it is often possible to map Monte Carlo ``dynamics'' onto Brownian dynamics (with physical units) by introducing an effective time scale and comparing, e.g., decorrelation times to known results \cite{weichsel_more_2016,whitelam_approximating_2011,levis_clustering_2014}. In Ref.~\onlinecite{weichsel_more_2016}, for example, this was achieved for a similar triangular mesh-based membrane model with fluctuating area under fixed tension. We proceed here under the assumption that such a mapping is possible for our system. 

The discussion that follows in this section focuses primarily on the dependence of the net flux of surface particles (vertices), $\Delta N(t)$, on nonequilibrium driving. The flux of surface particles is directly proportional to the change in total surface area $\Delta A(t)$ because the average number of vertices per unit area, $\rho_0$, is approximately constant due to the area constraint energy term in Eq.~\ref{eq:energy}. Equivalently, we can write $A(t) \approx \rho_0^{-1} N(t)$. 

Fig.\ \ref{fig:fig_2}a displays ensembles of trajectories of the net flux of surface particles $\Delta N(t) = N(t) - N(0)$ as functions of time $t$ for three values of the excess chemical potential, $\Delta\mu = \mu - \mu_\text{eq}$. Significant fluctuations in the net particle flux $\Delta N(t)$ occur both along individual trajectories and across the ensemble of trajectories. For a given time $t$, the first moment of the distribution of net fluxes is the mean flux $\langle \Delta N(t)\rangle $, in which $\langle \cdot \rangle$ denotes an average over all $n_\mathrm{samples}$ trajectories, and the second central moment is the variance  $\mathrm{Var}(\Delta N(t)) \equiv \langle (\Delta N(t))^2 \rangle - \langle \Delta N(t)\rangle^2 $. 

At the equilibrium chemical potential $\mu_\mathrm{eq}$, the long-time mean flux vanishes, i.e., $\lim_{t\to\infty} \langle \Delta N(t)\rangle = 0$. For the leftmost panel in Fig.\ \ref{fig:fig_2}a, in which the deviation from equilibrium is small ($\Delta\mu \approx 0.2$), the ensemble-averaged net flux $\langle \Delta N(t)\rangle$ slowly increases with an approximately linear dependence on time. A typical final vesicle configuration (shown in the upper half of the panel) remains quasispherical, with relatively small curvature fluctuations reminiscent of an equilibrated vesicle, albeit with a slightly larger average radius than the initial configuration. In the middle panel of Fig.\ \ref{fig:fig_2}a, in which $\Delta \mu \approx 2.8$, the deviation from equilibrium is more substantial; here, $\Delta N(t)$ increases more rapidly with time, and the final configuration exhibits somewhat larger-amplitude shape fluctuations. After an initial transient regime, the system appears to enter a steady growth regime in which $\Delta N(t)$ increases linearly with time. The occurrence of this initial transient regime is evidently related to $p_\mathrm{exchange}$, the rate at which particle exchange moves are attempted; here, $p_\mathrm{exchange}=1$, such that the number of addition and removal moves attempted during a sweep is equal to the number of particles at the start of the sweep. For a lower particle exchange attempt rate of $p_\mathrm{exchange}=0.01$, the initial transient regime vanishes, whereas the linear growth regime extends to at least $t= 5\times 10^4$ steps  (see Fig. S1 in the Supplemental Material \cite{SI}).

For far-from-equilibrium values of the excess chemical potential $\Delta\mu$ (rightmost panel of Fig.\ \ref{fig:fig_2}a, with $\Delta \mu \approx 7.5 k_\mathrm{B}T$), we find that  $\langle\Delta N(t)\rangle$ grows nonlinearly at long times, with vesicle shapes exhibiting large-amplitude undulations that appear to reflect localized mechanical instabilities, reminiscent of the shapes observed in the experiments of Ref.~\onlinecite{solon_negative_2006}. The $\Delta\mu$-dependence of this behavior is summarized in Fig.\ \ref{fig:fig_2}b, in which periodic snapshots of configurations from sample trajectories at varying $\Delta\mu$ are shown. We observe qualitatively consistent behavior in two-dimensional vesicles, in which increasing $\Delta\mu$ likewise leads to a nonlinear growth regime with large-amplitude radial undulations (see Figs. S3 and S6 in the Supplemental Material \cite{SI}). Clearly, increasing the excess chemical potential $\Delta \mu$ induces a crossover between stable and unstable growth regimes with distinct morphological characteristics. We can rationalize the onset of this morphological transition as a consequence of the renormalization of effective mechanical properties by nonequilibrium driving.

\subsection{Renormalization of the effective elastic properties by nonequilibrium driving}\label{sec:renormalized_parameters}

The enhancement of curvature fluctuations with nonequilibrium driving can be interpreted as changes in the effective elastic properties of the membrane \cite{prost_shape_1996,manneville_activity_1999,girard_passive_2005,solon_negative_2006, turlier_equilibrium_2016,turlier_unveiling_2019}. The shape of the fluctuating  quasispherical vesicle is described by a superposition of spherical harmonic modes
\begin{equation}
r(\theta, \phi, t) = R(t)\left(1 + \sum_{\ell \ge 2}^{\ell_\mathrm{max}}\sum_{m=-\ell}^{\ell}u _{\ell m}(t) Y_{\ell m }(\theta, \phi)\right),
\end{equation}
in which $R(t)$ and $u_{\ell m}(t)$ correspond to the average radius and the spherical harmonic mode amplitudes, respectively, at time $t$ \cite{seifert_concept_1995,seifert_fluid_1999}.  We treat the set of $n_\mathrm{samples}$ configurations at elapsed time  $t=\tau$ as a configurational ensemble. To simplify notation, we hereafter use $u_{\ell m}=u_{\ell m}(\tau)$. We consider the ensemble average of the squared amplitude of modes of degree $\ell$, given by $\langle |u_\ell|^2 \rangle$, in which $|u_\ell|^2= (2\ell+1)^{-1}\sum_m |u_{\ell m}|^2$.

We find that, unexpectedly, the ensemble-averaged spectrum $\langle |u_\ell|^2\rangle $ is well approximated by the predicted spectrum for an externally pressurized elastic shell, adapted from Refs.\ \onlinecite{paulose_fluctuating_2012,kosmrlj_statistical_2017, rochal_viscoelastic_2005}:
\begin{equation} \label{eq:spectrum}
\langle |u_\ell|^2\rangle = \frac{k_\mathrm{B} T}{2 A_\ell},
\end{equation}
in which the quantity $A_\ell$ is defined as
\begin{equation} \label{eq:Al}
      \begin{split}
   A_{\ell} = & \; \gamma R^2 \left(1 + \frac{\ell(\ell+1)}{2}\right) + \frac{Y R^2}{2}  \frac{(\ell + 2)(\ell - 1)}{\ell(\ell + 1) } \\
    & + \frac{\kappa}{2}  (\ell + 2)^2 (\ell - 1)^2 .
  \end{split}
\end{equation}
Here, $\gamma$ is the effective tension \footnote{Note that, in Refs. \onlinecite{paulose_fluctuating_2012,kosmrlj_statistical_2017}, the spectrum is written in terms of an externally applied pressure $P$, equivalent to a negative tension  $\gamma = -P R/2$. Here, we use a capital $P$ to distinguish this externally applied pressure from our osmotic pressure difference, $\Delta p$.}, $Y$ is the effective two-dimensional Young's modulus \footnote{The two-dimensional Young's modulus is  defined as $Y = 4 \mu' \left(\mu' + \lambda'\right)/(2 \mu' + \lambda')$, in which $\mu'$ and $\lambda'$ are the Lam\'e coefficients. Here, we assume in-plane incompressibility ($\lambda'\to\infty$). For finite $\lambda'$, however, the denominator of the Young's modulus term  in Eq. \ref{eq:Al} includes an additional contribution $-2\mu'/(2\mu'+\lambda')$. For $\lambda' = \mu'$, one obtains the Young's modulus term of Ref. \onlinecite{paulose_fluctuating_2012}}, and $\kappa$ is the effective bending rigidity.  

It is noteworthy that our measured spectra readily fit an elastic shell model, despite the fact that our Monte Carlo move set explicitly allows for edge flipping moves (see Appendix \ref{app:simulation_details_3D}) that continuously rearrange the mesh topology and should, in principle, endow the vesicle with in-plane fluidity  \cite{gompper_network_1997,noguchi_shape_2005,noguchi_dynamics_2005,sadeghi_particle-based_2018}. Possibly, this apparent elasticity emerges if the characteristic timescale for in-plane rearrangements (controlled in part by $p_\mathrm{flip}$) is longer than the timescale for particle exchange (controlled in part by $p_\mathrm{exchange}$). Alternatively, this could perhaps be a consequence of the direct suppression of in-plane rearrangement by particle exchange; for example, the sufficiently rapid addition of particles to the surface could lead to in-plane ``jamming'' such that particle rearrangements become energetically costly. This warrants more systematic investigation in a future study. 

For sufficiently large $\Delta\mu$, we find that the measured spectra develop a peak at a characteristic degree, $\ell^*$, corresponding to the spherical harmonic mode into which excess area is concentrated. The wavelength corresponding to the peak degree $\ell^*$, $\lambda^*= 2\pi R/\ell^*$
decreases as  the  excess chemical potential $\Delta\mu$ increases indicating that the vesicles become increasingly wrinkled (see Fig.\ S8a-b in the Supplemental Material \cite{SI}).

\begin{figure*}[htbp!]
   \centering
   \includegraphics[width=7in]{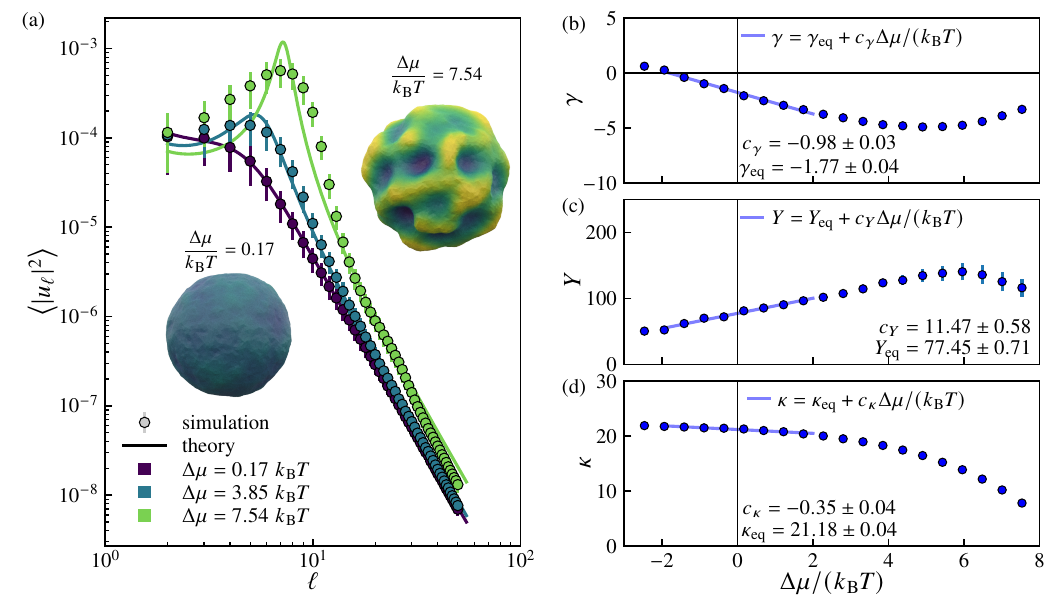}
   \caption{\textbf{Power spectra and renormalized mechanical properties of growing vesicles at various excess chemical potentials.} (a) Mean squared amplitude $\langle |u_\ell|^2 \rangle$ of spherical harmonic modes of degree $\ell$ for three different values of excess chemical potential $\Delta\mu$. Solid curves show fits to Eq.~\ref{eq:spectrum}, an effective elastic shell model with $\Delta\mu$-dependent renormalized tension $\gamma$ (in units of $k_\mathrm{B}T/\ell_0^2$), Young's modulus $Y$ (in units of $k_\mathrm{B}T/\ell_0^2$), and bending rigidity $\kappa$ (in units of $k_\mathrm{B} T$). Representative vesicle configurations illustrate the transition from quasispherical to highly deformed  morphologies with increasing $\Delta\mu$. (b-d) Variation of the renormalized parameters with $\Delta\mu/k_\mathrm{B}T$. The near-equilibrium regime ($\Delta\mu \approx 0$) shows approximately linear behavior characterized by slopes $c_\gamma$, $c_Y$, and $c_\kappa$. Here, the particle exchange attempt rate is $p_\mathrm{exchange}=1$, the osmotic pressure difference is $\Delta p = 0$, the cutoff time is $\tau=5000$ sweeps, and the number of samples is $n_\mathrm{samples} = 500$.}
   \label{fig:fig_4}
\end{figure*}

Fig.\ \ref{fig:fig_4}a shows the measured amplitude spectra $\langle| u_\ell |^2\rangle$ as a function of spherical harmonic degree $\ell$ for three increasing values of the excess chemical potential $\Delta \mu$. Fitting the spectra to  Eq.\ \ref{eq:spectrum} yields  a set of $\Delta\mu$-dependent effective mechanical properties $\gamma$, $Y$, and $\kappa$, which are shown in Fig.\ \ref{fig:fig_4}b-d. For $\Delta p = 0$ and small $\Delta\mu$, the effective mechanical properties vary linearly with $\Delta \mu/k_\mathrm{B} T$ over a regime extending to several $k_\mathrm{B} T$ beyond equilibrium. Here, vesicles settle into a state of fluctuating but steady growth, in which the extracted effective properties are insensitive to the elapsed time $\tau$. For the effective tension $\gamma$, the proportionality constant $c_\gamma$ is negative, such that increasing the excess chemical potential $\Delta\mu$ decreases the effective tension. The renormalization of the effective tension due to increasing $\Delta\mu$ is quite dramatic: for $\Delta\mu = 2k_\mathrm{B}T$, the effective tension $\gamma$ more than doubles the magnitude of the corresponding equilibrium value, $\gamma_\mathrm{eq}$. For values of $\Delta\mu$ above approximately $4 k_\mathrm{B} T$, the dependence of each property on $\Delta\mu$ is no longer linear: notably, the magnitudes of the effective tension and Young's modulus both apparently begin to decrease with increasing $\Delta\mu$. In this large-$\Delta\mu$ regime, vesicle shapes become highly nonspherical and vary considerably with time, as do the extracted effective properties. However, it should be noted that the theoretical spectrum (Eqs. \ref{eq:spectrum} and \ref{eq:Al}) fits the data poorly in this regime, implying that the effective parameters that we extract in this regime are no longer meaningful. In Fig. S5 in the Supplemental Material \cite{SI}, we present fluctuation spectra for two-dimensional vesicles with varying $\Delta \mu$. Fitting these spectra to Eq. \ref{eq:2d_helfrich} yields the effective tension $\gamma$ and bending rigidity $\kappa$. Consistent with our three-dimensional simulations, we find that increasing $\Delta\mu$ leads to a reduction in both $\gamma$ and $\kappa$. 

\begin{figure}[htbp!]
   \centering
   \includegraphics[width=3.375in]{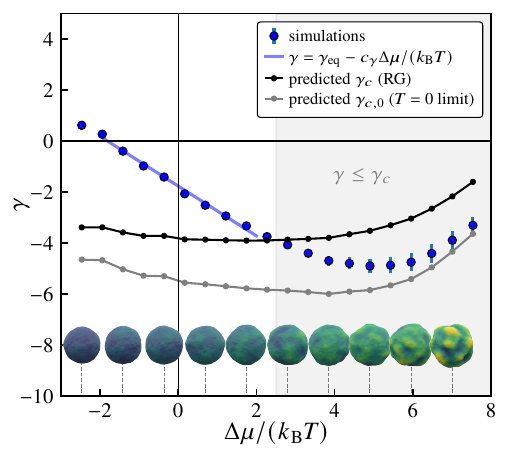}
    \caption{\textbf{Renormalized material parameters predict the onset of wrinkling.}  Variation of the  effective tension $\gamma$ with increasing excess chemical potential $\Delta\mu/k_\mathrm{B} T$. Blue points show values extracted from fitting vesicle shape fluctuation spectra to Eq.~\ref{eq:spectrum}. The gray curve shows the predicted zero-temperature critical buckling tension $\gamma_{c,0}=-(2/R)\sqrt{\kappa Y}$, while the black curve shows the temperature-dependent critical tension $\gamma_c=\gamma_{c,0}\Psi(\mathsf{ET})$ predicted by RG calculations from Ref.~\onlinecite{kosmrlj_statistical_2017}. The light blue line shows a linear fit in the near-equilibrium regime. The shaded region indicates where buckling is predicted to occur based on the RG calculations, with $\gamma < \gamma_c$. Also shown are representative final configurations corresponding to the values of $\Delta \mu$ indicated by the dashed lines. Here, $p_\mathrm{exchange}=1$, $\Delta p = 0$, and $\tau=5000$ sweeps, and the number of samples is $n_\mathrm{samples} = 500$.}
   \label{fig:fig_5}
\end{figure}

We next address the physical mechanism of the observed wrinkling behavior by taking advantage of the effective equilibrium elastic shell description. At zero temperature, a spherical elastic shell is predicted to buckle at
a critical external pressure 
$P_{c,0} = 4 R^{-2}\sqrt{\kappa Y}$ \cite{paulose_fluctuating_2012}, or equivalently a negative effective tension $\gamma_{c,0} = -(2/R)\sqrt{\kappa Y}$. 
For shells at finite temperature, thermal fluctuations lead to a reduction in the critical buckling pressure \cite{paulose_fluctuating_2012,kosmrlj_statistical_2017}, which can be written as $P_c =  P_{c,0}\Psi\left(\mathsf{ET}\right)$. Here, $\Psi(x)$ is a monotonically decreasing scaling function  (see Fig.\ S9 in the Supplemental Material \cite{SI}) obtained in Ref.\ \onlinecite{kosmrlj_statistical_2017} using a renormalization group (RG) approach, and $\mathsf{ET}=(k_\mathrm{B} T/\kappa)\sqrt{(YR^2)/\kappa}$  \cite{kosmrlj_statistical_2017,agrawal_active_2023} is the (dimensionless) elastothermal number, which quantifies relative resistance to purely elastic and thermally induced deformations. By analogy, for our system, we expect the finite-temperature critical tension to exhibit the same dependence, i.e., $\gamma_c = \gamma_{c,0} \Psi(\mathsf{ET})$. It is important to note that the effective elastic constants are expected to depend on temperature \cite{kosmrlj_statistical_2017}. In calculating $\gamma_{c,0}$ and $\mathsf{ET}$ as functions of $\Delta\mu$ and $\Delta p$, we  assume that this temperature-dependent renormalization can be neglected, such that $\kappa(T) \approx \kappa(T=0)$ and $Y(T) \approx Y(T=0)$ at the temperature under consideration. For the bending rigidity, this assumption appears to be justified, as the measured effective equilibrium bending rigidity $\kappa_\mathrm{eq}\approx 21 k_\mathrm{B}T$ is very close to the specified microscopic value ($20 k_\mathrm{B} T$). Here, we are able to apply the renormalized critical buckling tension derived in Ref. \cite{kosmrlj_statistical_2017} because our measured spectra agree with the thermalized elastic shell model (Eqs. \ref{eq:spectrum} and \ref{eq:Al}). But, as we discussed earlier, this apparent elasticity may be a consequence of the parameter range under consideration (in particular, the choice of the particle exchange attempt probability $p_\mathrm{exchange}$).  Because the temperature-dependent renormalization of elastic constants for elastic shells differs from that of fluid vesicles \cite{leibler_equilibrium_2004}, a different critical tension would presumably be appropriate in the fluid regime. We leave this for future investigation.

Using the effective parameters $\kappa$ and $Y$ as a function of $\Delta\mu$, we calculate the $\Delta\mu$-dependent elastothermal number $\mathsf{ET}$ to obtain the predicted finite-temperature critical tension, $\gamma_c$, as a function of $\Delta \mu$ (see Fig.\ \ref{fig:fig_5}).  The finite-temperature buckling theory predicts buckling when $\gamma\le\gamma_c$. As the excess chemical potential $\Delta\mu$ increases from $0$, we see that the measured tension $\gamma$ becomes increasingly negative, eventually crossing the critical value $\gamma_c$. Representative final configurations (for $t=\tau$, at the end of the trajectory) are also shown for various values of $\Delta\mu$ on either side of the predicted crossover between non-buckled and buckled regimes, using the same coloring scale as in Fig.\ \ref{fig:fig_2}b. The regime over which the simulated nonequilibrium configurations exhibit wrinkling is roughly consistent with the predictions of the finite-temperature  elastic shell buckling model in the absence of driving. Notably, the measured effective tension does not cross the zero-temperature tension threshold. Since the observed onset of wrinkling is associated with the unstable growth dynamics described in the previous section, this implies that thermal fluctuations may play a role in giving rise to unstable growth. For shrinking vesicles with negative $\Delta\mu$, the spherical state is stabilized: as $\Delta\mu$ becomes increasingly negative,  the effective surface tension $\gamma$ is increased relative to the equilibrium value $\gamma_\mathrm{eq}$, eventually becoming positive.

\begin{figure}[htbp!]
   \centering
   \includegraphics[width=3.375in]{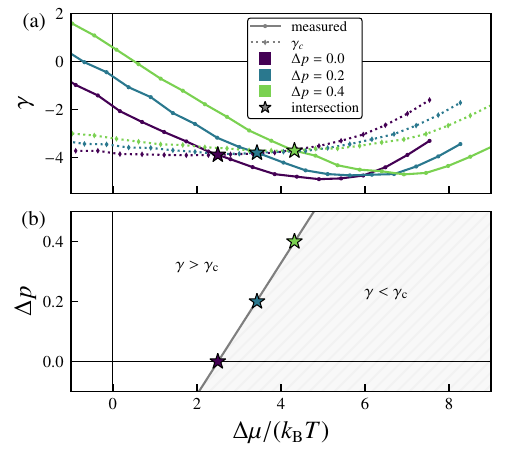}
   \caption{\textbf{Relationship between applied pressure, chemical potential, and vesicle stability.}  (a) Variation of the effective tension $\gamma$ with excess chemical potential $\Delta\mu/k_\mathrm{B}T$ for different values of osmotic pressure difference $\Delta p$. The measured effective tensions are compared with the predicted finite-temperature critical tension $\gamma_c$. The intersections (stars) indicate predicted morphological transition points, above which $\gamma < \gamma_c$. (b) Phase diagram in $\Delta p$-$\Delta\mu$ space. The region where $\gamma > \gamma_c$ corresponds to stable quasispherical growth, while $\gamma < \gamma_c$ indicates unstable growth with persistent out-of-plane deformations. The solid line represents a linear fit. Here, the excess chemical potential is computed relative to the equilibrium chemical potential at finite pressure difference: $\Delta\mu = \mu - \mu_\mathrm{eq}\big|_{\Delta p}$. For these data, $p_\mathrm{exchange}=1$, $\tau=5000$ sweeps, and the number of samples is $n_\mathrm{samples} = 500$.}
   \label{fig:phase_diagram_delta_mu_vs_delta_p}
\end{figure}

We next consider the more general case of nonzero osmotic pressure $\Delta p$. As $\Delta p$ increases, the equilibrium chemical potential $\mu_\mathrm{eq}$ decreases, while the equilibrium effective tension $\gamma_\mathrm{eq}$ increases (see Fig.\ S10a-b in the Supplemental Material \cite{SI}). In the linear regime, the effective tension $\gamma$ varies with the chemical potential $\mu$ and osmotic pressure $\Delta p$  as
$\gamma - \gamma_\mathrm{eq}\big|_{\Delta p} = c_\gamma (k_\mathrm{B}T)^{-1} \left(\mu - \mu_\mathrm{eq}\big|_{\Delta p}\right)$.  Fig.\ \ref{fig:phase_diagram_delta_mu_vs_delta_p}a extends the analysis of Fig. \ref{fig:fig_5} to different $\Delta p$ values and shows that increasing the osmotic pressure shifts the onset of the morphological transition rightward (to larger $\Delta\mu)$. As one would intuitively expect, if the osmotic pressure difference $\Delta p$ (for which positive values indicate a net positive internal pressure) is increased, a greater excess chemical potential $\Delta \mu$ is required to induce wrinkling behavior (see Fig. \ref{fig:phase_diagram_delta_mu_vs_delta_p}b). We see qualitatively consistent behavior in two-dimensional vesicles, in which increasing the osmotic pressure $\Delta p$ at fixed $\Delta\mu$ leads to reduced radial undulations and comparatively stable growth dynamics (see Figs. S4 and S7 in the Supplemental Material \cite{SI}).  While we have considered the case of fixed osmotic pressure difference $\Delta p$ here for simplicity, it is important to note that maintaining a fixed $\Delta p$ as the vesicle volume changes necessitates a mechanism for maintaining a constant internal solute concentration. In general, and especially in living cells, one would expect $\Delta p$ to be a dynamic, volume-dependent driving force.

Although we have shown that the substantial morphological changes that we observe can be explained as a consequence of renormalized mechanical properties, linking these changes to the underlying driving forces and predicting the dynamics remains challenging. In the following section, we outline a strategy based on stochastic thermodynamics to infer these forces and develop a simplified growth law for the system.

\begin{figure}[htbp!]
\centering
\includegraphics[width=3.375in]{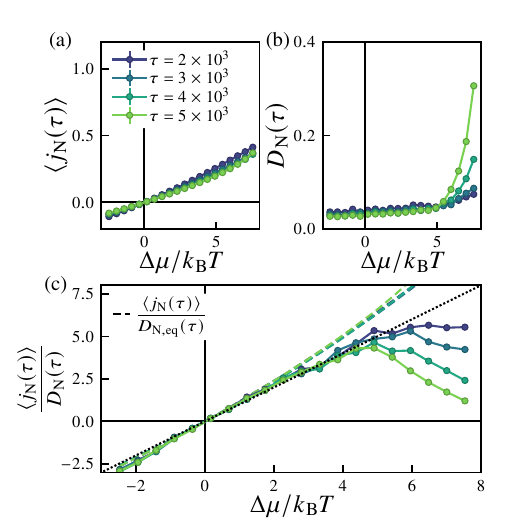}
\caption{\textbf{Dependence of the average particle flux, diffusivity, and their ratio on the excess chemical potential.} Shown are (a)  the average current $\langle j_\mathrm{N} (\tau)\rangle = \langle \Delta N(\tau)\rangle/\tau$, (b) the  diffusivity $D_\mathrm{N}(\tau) = \mathrm{Var}(\Delta N(\tau))/(2\tau)$, and (c) the ratio $\langle j_\mathrm{N}(\tau)\rangle/D_\mathrm{N}(\tau)$  as a function of the excess chemical potential $\Delta\mu$ for varying cutoff times $\tau$. In (c), the dashed lines show the same ratio if the denominator is replaced with the equilibrium diffusivity $D_\mathrm{N,eq} = D(\Delta\mu = 0)$, and the diagonal dotted line corresponds to the linear response prediction of Eq \ref{eq:fluctuation_dissipation}. For these simulations, the particle reservoir exchange attempt rate is $p_\mathrm{exchange} = 1$, the imposed osmotic pressure difference is $\Delta p = 0$, and the number of samples is $n_\mathrm{samples}=500$. }\label{fig:fluxes_etc}
\end{figure}

\subsection{A low-dimensional growth law inferred from stochastic thermodynamics}

Quantitatively predicting the shape dynamics of vesicles in far-from-equilibrium regimes remains an open problem with major implications for our understanding of both modern cell biology and the origin of life \cite{szostak_synthesizing_2001,ruiz-herrero_dynamics_2019, zhang_passive_2023}. 
Existing theoretical approaches based on linear irreversible thermodynamics (LIT) have made progress in developing covariant constitutive equations for membrane dynamics \cite{salbreux_mechanics_2017, sahu_irreversible_2017} but these formulations involve complex tensorial quantities and numerous unknown coupling coefficients, complicating their application to real systems. Moreover, the applicability of such frameworks \textit{far from equilibrium}, to processes such as growth-driven shape changes in vesicles, remains uncertain. We seek a minimal description that captures the essential physics of growing vesicles across regimes while remaining tractable for both theoretical analysis and experimental validation.

In this section, we build on our previous theoretical work \cite{nguyen_design_2016,nguyen_dissipation_2019,qiu_strong_2021} and use data generated from our nonequilibrium simulations to infer a simple, low-dimensional evolution equation for key observed variables. This approach leverages stochastic thermodynamics \cite{peliti_stochastic_2021}, a mathematical framework connecting thermodynamics to stochastic calculus, to associate a thermodynamic entropy cost to the observed fluctuating trajectories; this entropy cost implies an upper bound on the fluctuations of any coarse-grained set of observables \cite{gingrich_dissipation_2016,gingrich_inferring_2017,dechant_current_2018,seifert_stochastic_2019,dechant_multidimensional_2019,horowitz_thermodynamic_2020,dechant_improving_2021,dieball_direct_2023}. A growing body of recent work has demonstrated the utility of the so called thermodynamic uncertainty relations and their extensions in revealing new insights about nonequilibrium systems, particularly in experimental systems where limited observables are available \cite{di_terlizzi_variance_2024}, in nonequilibrium systems where off-diagonal couplings (between multiple forces or fluxes) are present \cite{qiu_strong_2021,chatzittofi_thermodynamic_2024, chatzittofi_collective_2023}, and in the inference of driving forces \cite{chatzittofi_entropy_2024}. As the entropy produced in a nonequilibrium process represents a sum of products of fluxes and forces \cite{de_groot_non-equilibrium_1984}, discovering a set of observables that saturate the aforementioned bounds can enable us to propose a dynamical law relating the observed fluxes $\bm{J}(\tau)$ (or the corresponding time-averaged currents $\bm{j}(\tau) = \bm{J}(\tau)/\tau$)  to the underlying thermodynamic driving forces. Our previous work \cite{nguyen_design_2016,nguyen_organization_2021} demonstrated that this workflow can be justified in highly simplified contexts. However, whether a similar result holds for growing membranes or other similarly complex systems remains an open question, which we address here.

We first consider the dependence of the particle flux and its fluctuations on the strength of nonequilibrium driving, the excess chemical potential $\Delta\mu$. Fig.\ \ref{fig:fluxes_etc}a shows the ensemble-averaged particle current (proportional to the surface area growth rate) $\langle j_\mathrm{N}(\tau)\rangle\equiv \tau^{-1}\langle\Delta N(\tau)\rangle$ as a function of the excess chemical potential $\Delta\mu$ for several values of $\tau$. The fluctuations of the net particle flux are quantified by the diffusivity $D_\mathrm{N}(\tau)$, given by $D_\mathrm{N}(\tau) \equiv (2\tau)^{-1}\mathrm{Var}(\Delta N(\tau)) $,
shown in Fig.\ \ref{fig:fluxes_etc}b. The diffusivity depends weakly on $\Delta \mu$ and $\tau$ near equilibrium, whereas far from equilibrium (for $\Delta \mu \gtrsim 5 k_\mathrm{B} T$) we find that $D_\mathrm{N}(\tau)$ increases both with increasing $\tau$ and with increasing $\Delta \mu$.  This behavior notably occurs in the regime in which highly deformed morphologies are observed. In the near-equilibrium regime (for $\Delta\mu \lesssim 5 k_\mathrm{B} T$), we find that the ensemble mean particle current obeys a linear response relation of the form
\begin{equation} \label{eq:fluctuation_dissipation}
    \langle j_\mathrm{N}(\tau)\rangle \approx \frac{D_\mathrm{N}(\tau)}{k_\mathrm{B} T} \Delta \mu,
\end{equation}
as plotted in Fig.~\ref{fig:fluxes_etc}c. This is essentially an application of the fluctuation-dissipation theorem \cite{callen_irreversibility_1951,callen_theorem_1952,kubo_fluctuation-dissipation_1966}, which applies in the limit of small $\Delta\mu$.  This behavior is quite robust to changes in our simulation parameters: we observe the same behavior for systems  with lower particle exchange attempt rates (Fig. S1 in the Supplemental Material \cite{SI}), with finite pressure (Fig.\ S2), and for both non-pressurized and pressurized two-dimensional vesicles (Figs.\ S3 and S4, respectively).

\begin{figure}[htbp!]
   \centering
   \includegraphics[width=3.375in]{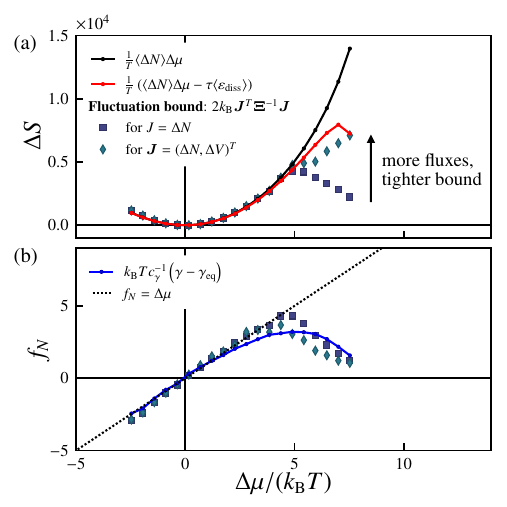}
   \caption{\textbf{Total entropy production and inferred driving force underlying material influx}. (a) Estimates and bounds for the total entropy production $\Delta S_{\text{tot}}$ as a function of the excess chemical potential $\Delta\mu/k_\mathrm{B}T$: Eq. \ref{eq:entropy_production} (black and red curves) as well as the fluctuation bound for a single current $J=\Delta N$ (circles) and for a two-dimensional current vector $\bm{J}=(\Delta N, \Delta V)^T$ (diamonds). (b) Inferred force $f_\mathrm{N}$ as a function of $\Delta\mu/k_\mathrm{B}T$ for different choices of current vectors $\bm{J}$, compared with the estimate proportional to the deviatoric  tension $k_\mathrm{B} T c_\gamma^{-1} (\gamma - \gamma_{\text{eq}})$ (blue curve). The dotted line corresponds to $f_\mathrm{N} = \Delta\mu/k_\mathrm{B} T$. Here, $p_\mathrm{exchange}=1$, $\Delta p = 0$, $\tau=5000$ sweeps, and the number of samples is $n_\mathrm{samples} = 500$.}
   \label{fig:fig_6}
\end{figure}

We next estimate the entropy production for this system following Ref.~\cite{nguyen_design_2016}. For a growing system, in the absence of any other fluxes, the entropy production is expected to be
\begin{equation} 
\label{eq:entropy_production}
    \Delta S = \frac{1}{T}\left(\Delta \mu \Delta N - \tau \langle\varepsilon_\mathrm{diss}\rangle\right) 
\end{equation}
in which $\langle \varepsilon_\mathrm{diss}\rangle$ is a relative entropy contribution, defined in Ref.~\onlinecite{nguyen_design_2016}, that provides a measure of how strongly the nonequilibrium configurational distribution deviates from equilibrium. We detail the calculation of $\langle\varepsilon_\mathrm{diss}\rangle$ in Sec.\ \ref{sec:relative_entropy}. The calculated values of $\Delta S$ are shown with and without the relative entropy contribution in Fig.\ \ref{fig:fig_6}. With a pressure difference, we expect $\Delta S = \frac{1}{T}\left(\Delta \mu \Delta N +\Delta p \Delta V - \tau \langle\varepsilon_\mathrm{diss}\rangle\right)$.  

Stochastic thermodynamics provides an alternative route for estimating bounds on the entropy production. Specifically, it provides methods for computing \textit{lower bounds} on the total entropy production using the statistical features of the observed fluxes. Motivated by Refs.~\onlinecite{dechant_multidimensional_2019,dechant_improving_2021}, we consider bounds of the form 
\begin{equation} \label{eq:mtur}
   \Delta S \ge 2 
   k_\mathrm{B} \langle \bm{J}\rangle ^T \bm{\Xi}^{-1} \langle \bm{J} \rangle
\end{equation}
for a vector of $d_f$ observed fluxes $\bm{J}(\tau)$ measured over an ensemble of trajectories of duration $\tau$, with covariance $\Xi_{ij} \equiv \mathrm{Cov}(J_i, J_j)$. The corresponding diffusivity matrix $\bm{D}(\tau)$ is given by $\bm{D}=(2\tau)^{-1} \bm{\Xi}$.  
A notable property of Eq.~\ref{eq:mtur} is that, as we increase the number of observed fluxes, $d_f$, the lower bound strictly increases, improving our estimate of the true entropy production $\Delta S$.

We next test, for our simulations, the effect of increasing the number of observables on the lower bound on $\Delta S$. In the regimes that we consider, the dominant flux is that of particles from the reservoir into the vesicle surface, i.e., $\Delta N$. When $\Delta N$ is the only observed flux,  Eq.\ \ref{eq:mtur} implies a lower bound on the entropy production of $\Delta S \ge 2 k_\mathrm{B} \langle \Delta N\rangle^2/\mathrm{Var}(\Delta N)$. Fig.~\ref{fig:fig_6}a shows this estimate as a function of $\Delta \mu$, and find that it saturates Eq. \ref{eq:entropy_production} in the near-equilibrium regime but deviates considerably in the far-from-equilibrium regime (for $\Delta\mu \gtrsim 5 k_\mathrm{B} T$). To improve the lower bound in this regime, it is clear that additional observables are required. We thus include the volume flux, $\Delta V$, such that the flux vector is $\bm{J} = (\Delta N,\;\Delta V)^T$. This considerably improves the performance of the lower bound in the far-from-equilibrium regime, in comparison to the single-observable estimate. 

If the observed fluxes are sufficiently informative, then the lower bound on the entropy production given by Eq.\ \ref{eq:mtur}) enables us to \textit{infer} the thermodynamic driving forces from the observed fluxes. Eq.~\ref{eq:mtur} allows us to estimate the vector of forces $\bm{f}=(f_\mathrm{N},\;f_\mathrm{V},\;...)^T$ as $\bm{f} \approx 2 k_\mathrm{B} T \bm{\Xi}^{-1} \langle \bm{J}\rangle $ or equivalently \begin{equation} \label{eq:inferred_force}
    \bm{f} \approx k_\mathrm{B} T \bm{D}^{-1} \langle \bm{j}\rangle,
\end{equation}
in which $\bm{j}(\tau)=\tau^{-1}\bm{J}(\tau)$ as specified earlier.  Fig.\ \ref{fig:fig_6}b shows, for $\Delta p = 0$, the inferred force vector conjugate to the surface particle flux, $f_\mathrm{N}$, computed using Eq.\ \ref{eq:inferred_force} with the same sets of observables as in Fig.~\ref{fig:fig_6}a. In the near-equilibrium regime ($\Delta\mu \lesssim 5 k_\mathrm{B}T$), we find that the inferred force $f_\mathrm{N}$ is  equivalent to the excess chemical potential $ \Delta\mu$, as expected. Given the linear relationship between the excess chemical potential and the change in the effective tension, $\gamma - \gamma_\mathrm{eq}$ in the near-equilibrium regime, we can relate the inferred force $f_\mathrm{N}$ and the deviatoric effective tension $\gamma-\gamma_\mathrm{eq}$ in this regime as $f_\mathrm{N} \approx k_\mathrm{B} T c_\gamma^{-1}(\gamma - \gamma_\mathrm{eq})$ with $c_\gamma = -0.98 \pm 0.03$ (see Fig.~\ref{fig:fig_6}b). Far from equilibrium ($\Delta\mu \gtrsim 5 k_\mathrm{B} T$), however, both the inferred force $f_\mathrm{N}$ computed via Eq.~\ref{eq:inferred_force} and the scaled deviatoric effective tension $k_\mathrm{B} T c_\gamma^{-1}(\gamma - \gamma_\mathrm{eq})$  decrease with increasing $\Delta \mu$.

\begin{figure}[htbp!]
    \centering
    \includegraphics[width=3.375in]{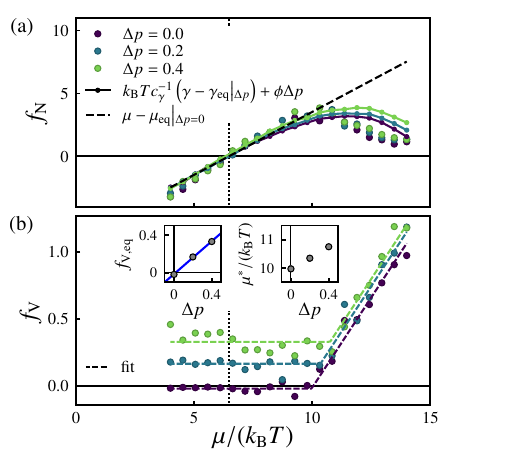}
    \caption{\textbf{Inferred thermodynamic forces as a function of chemical potential  under various imposed osmotic pressures.} (a) Inferred force $f_\mathrm{N}$ conjugate to the surface particle flux  $\Delta N$ (circles) and the effective tension-dependent estimate of Eq.~\ref{eq:fn}: $f_\mathrm{N}=k_\mathrm{B}T c_\gamma^{-1}\left(\gamma - \gamma_\mathrm{eq}\middle|_{\Delta p}\right)+\phi \Delta p$ (solid lines with small markers). The dashed line corresponds to $f_\mathrm{N} = \mu - \mu_\mathrm{eq}\big|_{\Delta p = 0}$, and the vertical dotted line corresponds to $\mu = \mu_\mathrm{eq}\big|_{\Delta p = 0}$. (b) Inferred force $f_\mathrm{V}$ conjugate to the volume flux  $\Delta V $. The dashed curves correspond to fits to Eq.~\ref{eq:fv}. The left inset shows the linear dependence of $f_{\mathrm{V,eq}} = f_\mathrm{V}(\mu = \mu_\mathrm{eq}\big|_{\Delta p=0})$ on $\Delta p$, with $f_\mathrm{V,eq}\approx c_p\Delta p$ in which $c_p = 0.87 \pm 0.03$ (blue line). The right inset shows the fit values of the threshold chemical potential $\mu^*$ beyond which $f_\mathrm{V}$ begins to increase with $\mu$. For these data, $p_\mathrm{exchange}=1$ and the number of samples is $n_\mathrm{samples}=500$.}
\label{fig:varying_delta_p}
\end{figure}

We next consider the more general behavior of the inferred forces under varying osmotic pressure $\Delta p$. Fig.~\ref{fig:varying_delta_p} shows the inferred forces $f_\mathrm{N}$ and $f_\mathrm{V}$ conjugate to the observed fluxes $\bm{J} = (\Delta N,\;\Delta V)^T$. Taking into account the $\Delta p$-dependence of the equilibrium chemical potential $\mu_\mathrm{eq}$ and equilibrium effective tension $\gamma_\mathrm{eq}$ (see Fig.\ S10 in the Supplemental Material \cite{SI}), the relationship between the $\Delta N$-conjugate thermodynamic driving force $f_\mathrm{N}$ and the pressure-dependent deviatoric effective tension $\left.\gamma - \gamma_\mathrm{eq}\middle|_{\Delta p}\right.$ can be written as 
\begin{equation} \label{eq:fn}
    f_\mathrm{N} \approx k_\mathrm{B} T c_\gamma^{-1} (\gamma - \gamma_\mathrm{eq}\big|_{\Delta p}) + \phi \Delta p.
\end{equation}
In the near-equilibrium regime (for $\Delta \mu \lesssim 5 k_\mathrm{B} T$), the driving force conjugate to the volume flux, $f_\mathrm{V}$, is independent of $\mu$ but proportional to $\Delta p$ with $f_\mathrm{V} \approx c_p \Delta p$ in which $c_p = 0.87\pm 0.03$ (see Fig. \ref{fig:varying_delta_p}b inset). For stronger driving ($\Delta \mu \gtrsim 5 k_\mathrm{B} T$), $f_\mathrm{V}$ begins to increase linearly with $\mu$, with approximately the same slope for all $\Delta p$. We can capture the behavior of the $\Delta V$-conjugate thermodynamic driving force $f_\mathrm{V}$ in both with the heuristic form
\begin{equation} \label{eq:fv}
    f_\mathrm{V} \approx c_p \Delta p + \zeta \left(\mu - \mu^*\big|_{\Delta p}\right)\Theta\left(\mu - \mu^*\big|_{\Delta p}\right),    
\end{equation}
plotted in Fig.~\ref{fig:varying_delta_p}b. Here, $\zeta=0.27\pm0.01$ is the slope, $\mu^*\big|_{\Delta p}$ is the $\Delta p$-dependent threshold chemical potential at which $f_\mathrm{V}$ begins to increase with $\mu$, and $\Theta$ is the Heaviside step function, defined as $\Theta(x)=0$ for $x<0$ and $\Theta(x)=1$ for $x\ge0$. The fit values of $\mu^*\big|_{\Delta p}$ are plotted in the inset.

Having demonstrated that the inferred thermodynamic driving forces are connected to physically interpretable quantities (the osmotic pressure and effective tension), we can in principle make predictions about the dynamics of our system---e.g., how the surface particle flux and volume flux vary with changes in tension. If we know the functional form of the forces $\bm{f}$, then we can compute the fluxes as $\langle \bm{J}\rangle \approx (2 k_\mathrm{B} T)^{-1}  \bm{\Xi} \bm{f}$,
or equivalently $\langle\bm{j}\rangle \approx (k_\mathrm{B} T)^{-1}\bm{D}\bm{f}$ \footnote{The diffusion matrix $\bm{D}/(k_\mathrm{B}T)$ here is, in the linear regime, equivalent to the matrix of phenomenological coefficients relating the forces and fluxes in the LIT approach, which one obtains by directly varying the forces. In Appendix \ref{sec:LIT}, we demonstrate that the phenomenological matrix satisfies Onsager's reciprocal relations (see Fig.\ \ref{fig:onsager}). }. This implies that we can write the average rates of change of the surface particle number $N$ and volume $V$ as
\begin{equation} \label{eq:eom}
        \begin{pmatrix} \langle \dot{N}\rangle \\ \langle \dot{V} \rangle\end{pmatrix}  \approx \frac{1}{k_\mathrm{B} T} \begin{pmatrix} D_\mathrm{N} & D_\mathrm{NV} \\ D_\mathrm{NV} & D_\mathrm{V}\end{pmatrix} \begin{pmatrix} f_\mathrm{N} \\ f_\mathrm{V} \end{pmatrix}
\end{equation}
in which we have defined $\dot{N}\equiv j_\mathrm{N}(\tau)=\Delta N(\tau)/\tau$ and $\dot{V}\equiv j_\mathrm{V}(\tau)=\Delta V(\tau)/\tau$ for simplicity.
Here, the components of the diffusion matrix $\bm{D}$ can be readily obtained from the fluctuations of observed trajectories, and we can insert the expressions for the thermodynamic driving forces $f_\mathrm{N}$ and $f_\mathrm{V}$ (as functions of $\Delta p$ and $\gamma - \gamma_\mathrm{eq}$) that we determined earlier (Eqs.~\ref{eq:fn} and~\ref{eq:fv}).
Note that, because the surface particle number and area are related as $N \approx \rho_0 A$, Eq.~\ref{eq:eom} can be readily converted into a shape equation relating changes in surface area and volume. This represents a simple ``equation of motion'' that describes the dependence of two key shape parameters---the surface particle number (proportional to the surface area) and volume of a growing vesicle---on experimentally accessible control variables, i.e., the membrane tension $\gamma$ and osmotic pressure $\Delta p$.   In a design context, one could use this relationship to optimize conditions for uniform growth or for triggering shape transformations on demand. For example, one can straightforwardly obtain the diffusivity matrix from experimentally observed trajectories, and then invert the equation to determine the precise values of $\gamma$ and $\Delta p$ required to achieve a desired growth trajectory or morphological state.
When compared to the rigorous LIT-based  descriptions developed in Refs. \onlinecite{salbreux_mechanics_2017, sahu_irreversible_2017} , the low-dimensional growth law given by Eq.~\ref{eq:eom}  has the advantages of being simple but valid far from equilibrium.

Although we have restricted our discussion here to fluxes of the surface particle number $N$ and vesicle volume $V$, our approach is general and can be readily extended to other observables. For example, one could include the flux of elastic energy into each spherical harmonic mode ($\ell = 2, 3, ..., \ell_\mathrm{max})$, which would provide information about which shape modes capture the most chemical work. Another possibility would be the excess area (or the related reduced volume $\nu$), which would more directly isolate shape changes than the currently used $N$ (which is proportional to the total surface area). Further, if explicit dynamics were simulated, one could consider the viscous dissipation associated with in-plane rearrangements as a separate flux. Finally, a straightforward option compatible with our current Monte Carlo framework would be to treat, as a flux, the medium entropy production (housekeeping heat). In general, including additional fluxes would tighten the fluctuation bound on the total entropy production and potentially provide a more informative low-dimensional ``equation of motion.''

\section{Conclusion}

Here, we combined Monte Carlo simulations with tools from statistical physics and stochastic thermodynamics to investigate the nonequilibrium growth of vesicles. In this system, an imposed excess chemical potential and osmotic pressure difference drive fluxes of surface particles and internal volume, respectively, into a fluctuating vesicle. We find that the shape fluctuation spectra of our growing vesicles are consistent with theoretical predictions for elastic shells subjected to an imposed in-plane tension. Within this description, we extract renormalized effective elastic parameters as a function of the excess chemical potential. Our results demonstrate that nonequilibrium driving leads to a renormalization of the effective mechanical properties---the effective tension, Young's modulus, and bending rigidity---of the vesicle surface, as suggested by recent theoretical work \cite{salbreux_mechanics_2017,sahu_irreversible_2017,bacher_computational_2019,bacher_three-dimensional_2021,torres-sanchez_interacting_2022}. Furthermore, we show that these changes in effective mechanical properties predict the onset of an instability reminiscent to the buckling instability of elastic shells. Specifically, as the driving strength increases, the effective tension becomes sufficiently negative to cross a theoretically predicted critical buckling threshold. Our simulations indicate that the onset of this instability  corresponds to a  transition from a stable growth regime with quasispherical vesicle configurations to an unstable growth regime with highly deformed morphologies exhibiting wrinkling. This behavior is reminiscent of chemically driven mechanical instabilities observed in elastic shells with oscillatory mechanochemical coupling \cite{li_chemically_2021}. These findings provide a quantitative link between the mechanical influence of nonequilibrium driving forces and macroscopic shape transformations.  The identification of a nonequilibrium wrinkling threshold governed by renormalized mechanical properties provides a quantitative design principle for controlling morphological transitions in biological contexts and in synthetic vesicles.

We leveraged ideas from stochastic thermodynamics \cite{dechant_multidimensional_2019,peliti_stochastic_2021} to infer a low-dimensional growth law for vesicles, in which the inferred thermodynamic driving forces are quantitatively related to the renormalized mechanical properties. This is a potentially powerful idea, as it provides a framework for quantitatively predicting the dynamical consequences of  changes in membrane tension in more general contexts, e.g., for situations in which the tension is modified due to the ATP-mediated activity of cytoskeletal filaments or membrane-embedded proteins \cite{ramaswamy_physics_2001}.

This framework could find natural application in biological systems where growth-driven mechanical instabilities drive cellular processes. Notably, the wrinkled vesicle morphologies that we observe are reminiscent of shapes seen in growing protocell models \cite{kindt_bulk_2020} and L-form bacteria \cite{briers_how_2012,errington_l-form_2013,mercier_excess_2013, koonin_evolution_2013}.  Although we have focused on growth, the analysis that we introduced can be readily applied to other processes---e.g., the dynamics of membranes driven by the polymerization of actin filaments \cite{weichsel_more_2016}, by active fusion and fission processes, \cite{rautu_active_2024}, or otherwise driven by the action of internal active matter \cite{paoluzzi_shape_2016,tian_anomalous_2017,wang_shape_2019,li_shape_2019,takatori_active_2020,vutukuri_active_2020,kokot_spontaneous_2022,diaz_active_2024,sciortino_active_2025}. Such processes are of great importance in a wide variety of cellular functions such as motility, division, and organelle formation.  Our inferred growth law provides a template for modeling the dynamics of these systems. Looking forward, it would be interesting to complement our TUR-based inference approach with model-free methods for estimating entropy production from observed trajectories \cite{ro_model-free_2022, li_measuring_2024, ji_identifying_2025}. We hope that our work will provide foundations for the quantitative modeling of protocell dynamics and stimulate further interest into the applications of thermodynamic uncertainty relations and their extensions in revealing new insights about nonequilibrium systems.

\section{Acknowledgments}

SV acknowledges support from the National Institute of General Medical Sciences of the NIH under Award No.\ R35GM147400.  JLS acknowledges support from the Eric and Wendy Schmidt AI in Science Postdoctoral Fellowship, a Schmidt Sciences program. ARD acknowledges support from the National Science Foundation (NSF) under Award No.\ MCB-2201235.   SV and PV  also acknowledge support by the National Institute for Theory and Mathematics in Biology, funded by NSF (DMS-2235451) and the Simons Foundation (MP-TMPS-00005320). This work was completed in part with resources provided by the University of Chicago’s Research Computing Center.


%

\appendix

\counterwithin*{figure}{part}

\stepcounter{part}

\renewcommand{\thefigure}{A\arabic{figure}}
\renewcommand{\thetable}{A\arabic{table}}

\section{Simulation details} \label{app:simulation_details}

\subsection{Three dimensions}\label{app:simulation_details_3D}

\subsubsection{Model description}
For the 3D case (simulations of two-dimensional membranes embedded in three-dimensional Euclidean space), we model the membrane as a triangulated mesh consisting of $N$ vertices $v_i$ with positions $\bm{r} = \{ \bm{r}_1, \bm{r}_2, ... \bm{r}_N \}$ connected by a set of edges $e_{ij}$.  We compute geometric quantities using standard tools from discrete differential geometry \cite{meyer_discrete_2003,gompper_network_1997, nelson_statistical_2004}. Using the discrete Laplace-Beltrami operator $\Delta$,  we compute the discrete mean curvature $H_i$ of the mesh at the position of vertex $i$ as
\begin{equation} \label{eq:discrete_mean_curvature}
H_i = \bm{n}_i \cdot \left(\Delta \bm{r}\right)_i = \bm{n}_i \cdot \frac{1}{\sigma_i}\sum_{j(i)} \frac{\sigma_{ij}}{l_{ij}}(\bm{r}_i - \bm{r}_j),
\end{equation}
in which the sum is taken over all neighbors $j$ of vertex $i$. Here, $\bm{n}_i$ is the unit vector normal to the mesh at the position of vertex $i$,
\begin{equation}
\bm{n}_i = \frac{1}{z_i} \sum_{f(i)} \bm{n}_f,
\end{equation} 
where $z_i$ is the valence of vertex $i$, the sum is taken over the adjacent faces $f(i)$, and $\bm{n}_f$ is the unit normal to face $f$, defined as
\begin{equation}
\bm{n}_f = \frac{\bm{e}_1 \times \bm{e}_2}{|\bm{e}_1 \times \bm{e}_2|},
\end{equation}
in which the basis vectors $\bm{e}_1$ and $\bm{e}_2$ correspond to two of the face's edges, chosen such that $\bm{n}_f$ is directed toward the vesicle exterior. In Eq. \ref{eq:discrete_mean_curvature},  $\sigma_i$ is the area of the virtual dual cell of vertex $i$,
\begin{equation}
\sigma_i = \frac{1}{4}\sum_{j(i)}\sigma_{ij} l_{ij},
\end{equation}
in which $l_{ij} = \| \bm{r}_i - \bm{r}_j \|$ is the length of edge $ij$ and $\sigma_{ij}$ is the length of the corresponding edge in the dual mesh,
\begin{equation}
\sigma_{ij} = l_{ij}\left[\cot(\theta_1) + \cot(\theta_2)\right],
\end{equation}
in which $\theta_1$ and $\theta_2$ are the angles opposite edge $ij$.

The total energy of the system $E_\mathrm{tot}$ is given by
\begin{equation}
E_\mathrm{tot}=E_\mathrm{bend} + E_\mathrm{tether} + E_\mathrm{area}
\end{equation}
The bending energy $E_\mathrm{bend}$ is computed as 
\begin{equation} \label{eq:discrete_bending}
    E_\mathrm{bend} = \frac{\kappa}{2}\sum_i \sigma_i \left[H_i-H_0\right]^2,
\end{equation}
in which the sum is taken over all vertices $i$ and $H_0$ is the spontaneous curvature; for our simulations, $H_0=0$.In the continuum limit \cite{gompper_random_1996,gompper_network_1997}, this is equivalent to the Helfrich Hamiltonian \cite{helfrich_elastic_1973}
\begin{equation}
E_\mathrm{H} = \frac{\kappa}{2}\int_\mathcal{S} \left[ (H-H_0)^2 \right] dA,
\end{equation}
in which $\kappa$ is the bending rigidity, $H$ is the mean curvature, and the integral is taken over the total membrane area. As our closed vesicles remain topologically unchanged throughout the simulation, we neglect Gaussian curvature contributions.  We include tethering and area potentials defined as in Ref.\ \onlinecite{vutukuri_active_2020}. 
The tether energy $E_\mathrm{tether}$ is calculated as
\begin{equation}
E_\mathrm{tether} =\sum_\mathrm{edges}\left[E_\mathrm{att}(l_{ij}) + E_\mathrm{rep}(l_{ij})\right],
\end{equation}
in which the attractive part $E_\mathrm{att}(r)$ is given by
\begin{equation}
E_\mathrm{att}(r)=\begin{cases} k_T \dfrac{\exp(1/(r_{cA}-r))}{r_\mathrm{max}-r} \; \mathrm{if} \; r > r_{cA} \\ 0 \; \mathrm{if} \; r\le r_{cA}\end{cases}
\end{equation}
and the repulsive part $E_\mathrm{rep}(r)$ is given by 
\begin{equation}
E_\mathrm{rep}(r)=\begin{cases} k_T \dfrac{\exp(1/(r-r_{cR}))}{r-r_\mathrm{min}} \; \mathrm{if} \; r < r_{cR} \\ 0 \; \mathrm{if} \; r\ge r_{cR}\end{cases}.
\end{equation}
Here, $k_T$ is the tether stiffness, $r_\mathrm{min}$ and $r_\mathrm{max}$ are the minimum and maximum tether length, respectively, and $r_{cR}$ and $r_{cA}$ are the repulsive and attractive interaction cutoffs, respectively.
The area energy $E_\mathrm{area}$ is computed as
\begin{equation}
E_\mathrm{area} = \sum_{\mathrm{faces}} \frac{k_A}{2}\frac{(A_f - A_{f}')^2}{A_f'},\;\text{with}\;A_f' = \frac{A_\mathrm{tot,0}}{N_f},
\end{equation}
in which $k_A$ is the area stiffness, $A_f$ is the area of face $f$, $A_\mathrm{tot,0}$ is the target total area, and $N_f$ is the total number of faces. 

\subsubsection{Monte Carlo details} \label{app:monte_carlo_3D}

As described in the main text, We denote the instantaneous state of the system by $\Gamma = \left[\left\{\bm{r}\right\}, N, V\right]$, with which we associate a function $\Phi(\Gamma)\equiv\Phi(\Gamma,\mu,\Delta p)$ defined as
\begin{equation}
    \Phi(\Gamma) = E_\mathrm{tot}(\left\{\bm{r}\right\}) - \mu N - \Delta p V.
\end{equation}
Here, $N$ is the number of particles in the vesicle surface, $\left\{\bm{r}\right\}$ is the corresponding set of $N$ particle positions, and $V$ is the vesicle volume, also specified by $\left\{\bm{r}\right\}$. We evolve the system stochastically using the Monte Carlo method, with a move set that includes vertex translation, edge flipping, vertex addition ($N\to N+1$), and vertex removal ($N \to N-1$) moves. Proposed state transitions $\Gamma\to\Gamma'$ are accepted according to the Metropolis criterion \cite{metropolis_equation_1953} with probability $P_\mathrm{acc}(\Gamma\to\Gamma')$ given by
\begin{equation}
    P_\mathrm{acc}(\Gamma \to \Gamma') = \min\left[1, \frac{P_\mathrm{gen}(\Gamma'\to\Gamma)}{P_\mathrm{gen}(\Gamma\to\Gamma') }\exp\left(-\frac{\Delta \Phi(\Gamma\to\Gamma')}{k_\mathrm{B} T}\right)\right],
\end{equation}
in which $P_\mathrm{gen}(\Gamma \to \Gamma')$ is the probability of proposing the move $\Gamma \to \Gamma'$. 

In a \textit{vertex displacement} move, the position $\bm{r}_i$ of a selected vertex $v_i$ is moved to a random point within a sphere of radius $d_\mathrm{max}$ centered on $\bm{r}_i$, in which $d_\mathrm{max}$ is chosen such that the typical acceptance probability is approximately 0.5. 

\begin{figure}[htbp!]
   \centering
   \includegraphics[width=3.375in]{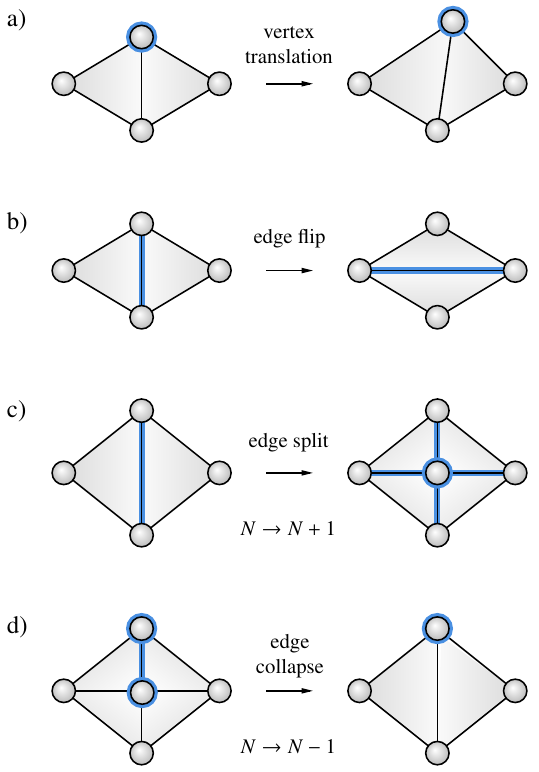}
 \caption{\label{fig:moveset} Monte Carlo move set for the triangulated mesh model for membranes in three dimensions. Circles and lines represent vertices and edges, respectively. }
 \end{figure}

We additionally include \textit{edge flip} moves, depicted in Fig.\ \ref{fig:moveset}b, which are commonly used in mesh-based membrane models to achieve fluidity \cite{noguchi_shape_2005,noguchi_dynamics_2005}. It has been shown in various prior studies that the edge-flipping frequency acts as a knob that controls the effective in-plane viscosity of the membrane. For example, Refs. \cite{noguchi_dynamics_2005,sadeghi_particle-based_2018} measured the effective in-viscosity as a function of bond-flipping frequency via gravity-driven in-plane Poiseuille flow. Here, for each edge, we attempt an edge flip in a given sweep with probability $p_\mathrm{flip}$.

Vertices are added via \textit{edge split} moves and removed via \textit{edge collapse} moves, depicted schematically in Fig.\ \ref{fig:moveset}c and d. In an edge split move, a new vertex is introduced at the midpoint of an existing edge, and new edges are added between the new vertex and the two vertices opposite the original edge, splitting each of the adjacent faces in two. In an edge collapse move, a selected vertex is merged with one of its neighbors by collapsing a randomly chosen adjacent edge, removing the two adjacent faces. Similar moves for the addition and removal of membrane material have been explored in prior Refs. \cite{weichsel_more_2016,sachin_krishnan_active_2022,rotskoff_robust_2018,okuda_continuum_2022}.

It is important to note that Monte Carlo simulations, as we have employed here, do not simulate the true dynamics of the system. However, one can reinterpret the transition probabilities $P(\Gamma\to\Gamma')$ as transition probabilities per unit time \cite{muller-krumbhaar_dynamic_1973,binder_introduction_1986,kremer_monte_1988,kremer_computer_2001}, although the underlying timescale involved has to be determined by other means.
Prior work has shown that one can make reasonable comparisons between Monte Carlo simulations and overdamped dynamics \cite{whitelam_approximating_2011,weichsel_more_2016}.

\subsection{Two dimensions}\label{app:simulation_details_2D}

\begin{figure}[htbp!]
\centering
\includegraphics[width=3.375in]{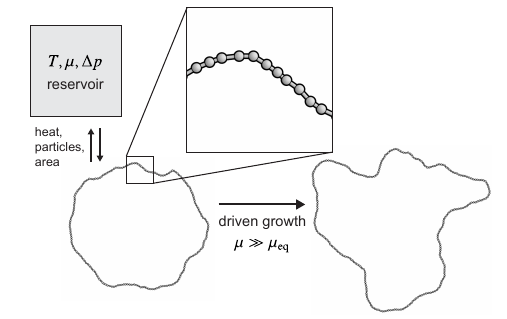}
\caption{Schematic of our Monte Carlo simulations for $d=2$, in which a fluctuating vesicle is modeled as a semiflexible polymer ring that exchanges heat, surface particles, and internal area with a reservoir characterized by temperature $T$, chemical potential $\mu$, and osmotic pressure difference $\Delta p$. For strongly nonequilibrium growth conditions ($\mu \gg \mu_\text{eq}$), we observe a morphological transition between a near-equilibrium regime in which the shapes of growing vesicles remain quasicircular and a far-from-equilibrium regime with persistent wrinkling.} \label{fig:2d_schematic}
\end{figure}

\begin{figure}[htbp!]
\centering
\includegraphics[width=3.375in]{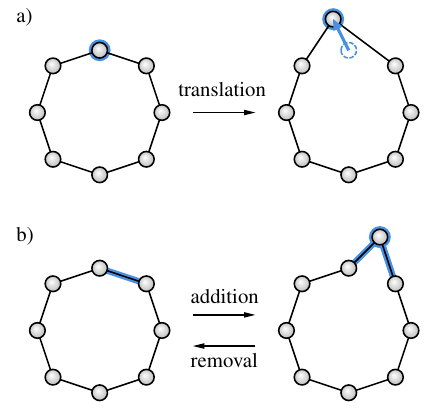}
\caption{Monte Carlo move set for the two-dimensional polymer ring model. (a) Vertex translation move, in which a randomly selected vertex is displaced randomly within a circle of radius $d_\mathrm{max}$. (b) Complementary particle addition and removal moves that allow the vesicle to grow or shrink by inserting or removing surface particles.} \label{fig:moveset_2d}
\end{figure}

\subsubsection{Model description}

In two dimensions, we model a vesicle as a closed semiflexible polymer (a ring polymer), which we couple to a reservoir with which it exchanges heat, surface particles, and internal area, as depicted in Fig.\ \ref{fig:2d_schematic}. As with the 3D model, we observe that above an equilibrium chemical potential $\mu_\mathrm{eq}$, then vesicles grow. Simple particle-based models have been used in the past for two-dimensional models for growing one-layered tissues \cite{drasdo_buckling_2000,drasdo_modeling_2000}.

The total potential energy $E_\mathrm{tot}$ of a vesicle configuration, in which the $i$th particle has position $\bm{r}_i$, is defined as
\begin{equation} \label{eq:2d_energy}
E_\mathrm{tot}=E_\mathrm{S} + E_\mathrm{B},
\end{equation}
in which $E_\mathrm{S}$ and $E_\mathrm{B}$  refer to the stretching and bending energies, respectively. The total stretching energy is defined as
\begin{equation}
E_\mathrm{S}=\sum_{ij}\frac{1}{2} \frac{k_S}{\ell_0}(|\bm{r}_{ij}|-\ell_0)^2,
\end{equation}
in which $k_S$ is the spring constant, $\ell_0$ is the rest length, $|\bm{r}_{ij}| = |\bm{r}_j - \bm{r}_i|$, and the sum is taken over all bonds $ij$. The total bending energy is defined as
\begin{equation}
E_\mathrm{B}=\sum_{ijk} \frac{1}{2}\frac{\kappa}{\ell_0}(\Delta\theta_{ijk})^2,
\end{equation}
in which $\theta_{ijk}$ is the angle between distinct pairs of adjacent bonds $ij$ and $jk$ and $\Delta\theta_{ijk} = \theta_{ijk} - \theta_0$. Note $\theta_i \approx \ell_0 c_i$, in which $c_i$ is the curvature at vertex $i$. We impose zero spontaneous curvature, such that $\theta_0 = 0$.

\subsubsection{Monte Carlo details}\label{app:monte_carlo_2D}

The general MC procedure for $d=2$ is virtually the same as for $d=3$. In this case, the function $\Phi(\bm{r}, N, V)$ is given by
\begin{equation} \label{eq:generalized_potential_2D}
    \Phi(\bm{r}, N, A) = E_\mathrm{tot}(\bm{r}) - \mu N - \Delta p A
\end{equation}
in which $N$ is the number of particles in the vesicle surface, $\bm{r}$ is the corresponding set of $N$ particle positions, and $A$ is the internal area, also specified by $\bm{r}$. 
The move set for $d=2$ includes vertex translation, addition, and removal moves, as shown schematically in Fig.~\ref{fig:2d_schematic}. For each sweep we attempt $N$ translation moves on randomly selected particles. For each translation move, the chosen particle is randomly displaced within a disk of radius $d_\mathrm{max}$ around its initial position. For translation moves, the forward and reverse move generation probabilities are symmetric: $P_\mathrm{gen}(\Gamma\to\Gamma')=P_\mathrm{gen}(\Gamma'\to\Gamma)$.

 For each sweep $N$ addition moves and $N$ removal moves with probability $p_\mathrm{exchange}$. For each addition or removal move, one of the $N$ particles is randomly selected. An addition move proposed for particle $i$ involves adding a new particle between particle $i$ and its counterclockwise neighbor, whereas a removal move involves removing $i$ and connecting its two neighbors.  In an addition move, the midpoint $\bm{r}_\mathrm{mid}$ between the chosen particle and its counterclockwise neighbor is identified, and a new candidate particle position is chosen randomly from a disk of area $a_\mathrm{add}$ centered on $\bm{r}_\mathrm{mid}$. The probability of generating a particular candidate particle addition move is thus
\begin{equation}
\begin{split}
P_\mathrm{gen}(\Gamma(N) \to \Gamma'(N + 1)) & = (N p_\mathrm{exchange})\frac{1}{a_\mathrm{add}N} \\ & = \frac{p_\mathrm{exchange}}{ a_\mathrm{add}}
\end{split}
\end{equation}
For the reverse (particle removal) move, the generation probability is
\begin{equation}
\begin{split}
P_\mathrm{gen}(\Gamma'(N + 1) \to \Gamma(N)) & = ((N + 1)p_\mathrm{exchange})\frac{1}{N+1} \\ & = p_\mathrm{exchange}
\end{split}
\end{equation}
In practice, we use $a_\mathrm{add} = \pi \ell_0^2/4$.

\subsubsection{Power spectrum }

We represent the shape of a 2D vesicle as a Fourier series, 
\begin{equation}
    r(\theta) = R \left[ 1 + \sum_{m=1}^{m_\mathrm{max}} \left(a_m \cos m \theta + b_m \sin m \theta \right) \right] .
\end{equation}
For an ensemble of shapes, the corresponding radial fluctuation spectrum is given by \cite{pecreaux_refined_2004,takatori_active_2020}
\begin{equation}
\langle |\hat{h}(q)|^2\rangle = \frac{\pi R_0^3}{2}\left(\langle |c_m|^2 \rangle - \langle |c_m|\rangle^2\right)
\end{equation}
in which $q = m/R_0$ is the wave number, $R_0 = \langle R\rangle $ is the ensemble-averaged radius, and $|c_m| = \sqrt{a_m^2 + b_m^2}$. We then fit the radial fluctuations to the theoretical form
\begin{equation} \label{eq:2d_helfrich}
\langle |\hat{h}(q)|^2\rangle = \frac{k_\mathrm{B} T}{\gamma q^2 + \kappa q^4}
\end{equation}
to extract effective values of the tension $\gamma$ and bending rigidity $\kappa$ (see Fig. S5 in the Supplemental Material \cite{SI}).

\subsection{Simulation parameters}

The parameters for our simulations in $d=3$ and $d=2$ are given in Tables \ref{table:simulation_parameters_d3} and \ref{table:simulation_parameters_d2}, respectively.

\begin{table}[htb!]
\begin{tabular}{lrr}
\toprule
\textbf{quantity}                  & \textbf{symbol}                  & \textbf{value} \\  
\midrule
thermal energy scale  & $k_\mathrm{B} T$  & $1$ \\
initial average tether length  & $\ell_0$   & $1$ \\
initial vesicle radius  & $R_0$   & $15\;\ell_0$ \\
initial vesicle surface area & $A_0$ & $\approx 4\pi R_0^2$ \\
initial number of vertices & $N_0$ & $\approx \left(4 \pi R_0^2\right)/\left(2\left(\frac{\sqrt{3}}{4} \ell_0^2\right)\right)$ \\
initial number of faces & $N_{f0}$ & $\equiv 2(N_0-2)$ \\
initial (target) area per face & $A_{f0}$ & $\equiv \frac{A_0}{N_{f0}}\approx \frac{\sqrt{3}}{4}\ell_0^2$ \\ 
bending rigidity  & $\kappa$  & $20 \; k_\mathrm{B}T$ \\    
tether stiffness  & $k_T$   & $100 \; k_\mathrm{B}T \ell_0$ \\    
local area stiffness  & $k_A$   & $100 \; k_\mathrm{B} T/\ell_0^2$ \\    
spontaneous curvature  & $H_0$   & $0 \; \ell_0^{-1}$ \\
flip attempt probability & $p_\mathrm{flip}$   & $0.3$ \\
max vertex displacement per step & $d_\mathrm{max}$ & $0.1 \; \ell_0$ \\
tether minimum length & $r_\mathrm{min}$ & $0.6 \; \ell_0$ \\
tether repulsive interaction cutoff & $r_{cR}$ & $0.8 \; \ell_0$\\
tether attractive interaction cutoff & $r_{cA}$ & $1.2 \; \ell_0$\\
tether maximum length & $r_\mathrm{max}$ & $1.4 \; \ell_0$ \\ 
\bottomrule
\end{tabular}
\caption{ \label{table:simulation_parameters_d3} Simulation parameters for $d=3$. }
\end{table}

\begin{table}[htb!]
\begin{tabular}{lrr}
\toprule
\textbf{quantity}                  & \textbf{symbol}                  & \textbf{value}  \\  
\midrule
thermal energy scale  & $k_\mathrm{B} T$  & $1$  \\
initial average tether length  & $\ell_0$   & $1$ \\
initial vesicle radius  & $R_0$   & $200\;\ell_0$ \\
initial vesicle perimeter & $P_0$ & $\approx 2\pi R_0$ \\
initial number of particles & $N_0$ & $\approx 2 \pi R_0/\ell_0$ \\
bending rigidity  & $\kappa$  & $20 \; k_\mathrm{B}T$ \\    
spring stiffness  & $k_S$   & $100 \; k_\mathrm{B}T \ell_0$ \\    
spontaneous curvature  & $H_0$   & $0 \; \ell_0^{-1}$ \\
max vertex displacement per MC step & $d_\mathrm{max}$ & $0.1 \; \ell_0$ \\
\bottomrule
\end{tabular}
\caption{ \label{table:simulation_parameters_d2} Simulation parameters for $d=2$.}
\end{table}

\FloatBarrier

\section{Relative entropy contribution} \label{sec:relative_entropy}

The thermodynamic reorganization cost $\langle \varepsilon_\mathrm{diss}\rangle$ \cite{nguyen_design_2016,nguyen_dissipation_2019}\ is defined as
\begin{equation}
\begin{split}
\langle \varepsilon_\mathrm{diss}\rangle & \equiv  \mathrm{D}_\mathrm{KL}\left[p_N || p_N^\mathrm{eq}\right] \\ & = \frac{\langle E_\mathrm{eq} - E_\mathrm{eff} \rangle_N - (F_\mathrm{eq} - F_\mathrm{eff})}{N}.
\end{split}
\end{equation}
We have
\begin{equation}
\langle E_\mathrm{eq}\rangle_N = \frac{k_\mathrm{B} T}{2} \sum_{\ell = 2}^{\ell_\mathrm{max}}(2\ell+1)\frac{A_{\ell,\mathrm{eq}}}{A_\ell},
\end{equation}
\begin{equation}
\langle E_\mathrm{eff}\rangle_N = \frac{k_\mathrm{B} T}{2} \sum_{\ell = 2}^{\ell_\mathrm{max}}(2\ell+1)\frac{A_{\ell}}{A_\ell} = \frac{k_\mathrm{B} T}{2}N,
\end{equation}
\begin{equation}
F_\mathrm{eq} = -\frac{k_\mathrm{B} T}{2}\sum_{2}^{\ell_\mathrm{max}}(2\ell+1) \ln \left(\frac{\pi k_\mathrm{B} T}{A_{\ell,\mathrm{eq}} }\right),
\end{equation}
and
\begin{equation}
F_\mathrm{eff} = -\frac{k_\mathrm{B} T}{2}\sum_{2}^{\ell_\mathrm{max}}(2\ell+1) \ln \left(\frac{\pi k_\mathrm{B} T}{A_{\ell} }\right),
\end{equation}
such that
\begin{equation}
\langle \varepsilon_\mathrm{diss}\rangle = \frac{1}{N}\frac{k_\mathrm{B} T}{2}\sum_{\ell = 2}^{\ell_\mathrm{max}}(2\ell+1)\left[\frac{A_{\ell,\mathrm{eq}}}{A_\ell}-1-\ln\frac{A_{\ell,\mathrm{eq}}}{A_\ell}\right].
\end{equation}
in which $\ell_\mathrm{max} = \sqrt{Nd}-1$. Here, $A_\ell$ and $A_{\ell,\mathrm{eq}}$ are defined by Eq. \ref{eq:Al} using the equilibrium and renormalized values of $\gamma$, $Y$, and $\kappa$.

\section{Phenomenological coefficients and Onsager reciprocity} \label{sec:LIT}

That we can write an expression for the dependence of the currents on driving forces (Eq.~\ref{eq:eom}) in the (linear) near-equilibrium regime is to be expected \cite{morris_dynamical_2010,morris_thermodynamics_2011}. It is well known that, within the framework of linear irreversible thermodynamics, we can write each current $j_i$ as a linear combination of the conjugate thermodynamic forces $f_j$, as
\begin{equation} \label{LIT_current}
j_i = \sum_{j} L_{ij} \, f_j,
\end{equation}
in which $\bm{L}$  is the phenomenological coupling matrix.  $\bm{L}$ is expected to be symmetric \cite{onsager_reciprocal_1931, onsager_reciprocal_1931-1,onsager_fluctuations_1953}, with $L_{ij} = L_{ji}$, and also positive semidefinite, with nonnegative diagonal entries $L_{ii}\ge0$ and a nonnegative determinant $\mathrm{det}(\bm{L})\ge 0$. 
In the linear regime, we should observe $L_{ij} = D_{ij}/k_\mathrm{B} T$. Note that the diffusion matrix $\bm{D}$ is \textit{by definition} symmetric and positive semidefinite, while these properties for $\bm{L}$ have to be checked by direct determination of the phenomenological coefficients, as $L_{ij} = \partial j_i/\partial f_j$. 

\begin{figure}[htbp!]
   \centering
   \includegraphics[width=3.375in]{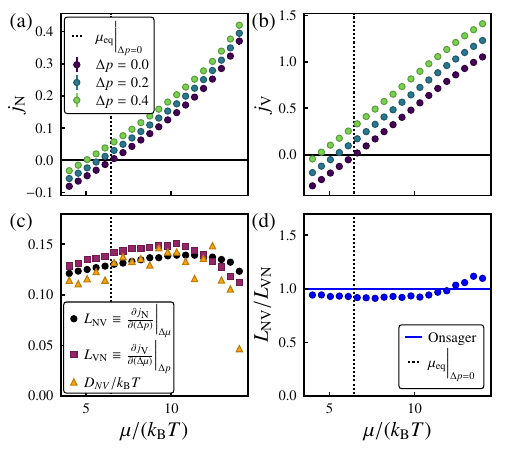}
   \caption{ \textbf{Symmetry in the nonequilibrium coupling coefficients}. For a given $\mu$, the  (a) ensemble-averaged surface growth rate $\langle \dot{N}\rangle $ and (b) ensemble-averaged volume growth rate $\langle \dot{V}\rangle$ both increase with increasing $\Delta p$.  The dotted line indicates the equilibrium chemical potential $\mu_{\text{eq}}$ for $\Delta p=0$. (b) The measured cross-coupling coefficients $L_\mathrm{NV}$ and $L_\mathrm{VN}$, agree well with one another, and with the corresponding cross-diffusion coefficient $D_\mathrm{NV}$. Note that $D_\mathrm{NV}=D_\mathrm{VN}$ by definition, due to the symmetry of the covariance matrix.  (d) The ratio $L_\mathrm{NV}/L_\mathrm{VN}$ is near unity, corresponding to the Onsager reciprocity relation (blue line). This holds true even far from the equilibrium chemical potential, indicated by the dotted black line. For these data, $p_\mathrm{exchange}=1$ and the number of samples is $n_\mathrm{samples} = 500$.}
   \label{fig:onsager}
\end{figure}

For our problem, we can test a version of the Onsager reciprocity relations by probing whether the following holds near and far from equilibrium, 
\begin{equation}
\frac{\partial \langle \dot{N}\rangle}{\partial (\Delta p)}\bigg|_{\Delta \mu} = \frac{\partial \langle \dot{V}\rangle}{\partial (\Delta \mu)}\bigg|_{\Delta p}
\end{equation}
That is, for systems near equilibrium, the dependence of the surface growth rate on the osmotic pressure, for fixed chemical potential, is identical to the dependence of the volume growth rate on the chemical potential, for fixed osmotic pressure. Fig.~\ref{fig:onsager}a shows both $L_\mathrm{NV}$ and $L_\mathrm{VN}$ as well as $D_\mathrm{NV}$, as a function of the imposed chemical potential $\mu$. We observe reasonable agreement between all three quantities over nearly the entire range of $\Delta\mu$ tested. To more directly demonstrate the near symmetry of the measured coupling coefficients, the ratio $L_\mathrm{NV}/L_\mathrm{VN}$ is shown in Fig.~\ref{fig:onsager}b.

\FloatBarrier
\end{document}


\hbadness=100000
\hfuzz=10000pt

\title{Supplemental Material for ``Renormalized mechanics and stochastic thermodynamics of growing vesicles''}

\author{Jordan L.\ Shivers}
\affiliation{The James Franck Institute, University of Chicago, Chicago, Illinois USA}
\affiliation{Department of Chemistry, University of Chicago, Chicago, Illinois USA}
\author{Michael Nguyen}
\affiliation{The James Franck Institute, University of Chicago, Chicago, Illinois USA}
\affiliation{Department of Chemistry, University of Chicago, Chicago, Illinois USA}
\author{Aaron R. Dinner}
\affiliation{The James Franck Institute, University of Chicago, Chicago, Illinois USA}
\affiliation{Department of Chemistry, University of Chicago, Chicago, Illinois USA}
\author{Petia M. Vlahovska}
\affiliation{Engineering Sciences and Applied Mathematics, Northwestern University, Evanston, Illinois USA}
\author{Suriyanarayanan Vaikuntanathan}
\affiliation{The James Franck Institute, University of Chicago, Chicago, Illinois USA}
\affiliation{Department of Chemistry, University of Chicago, Chicago, Illinois USA}

\maketitle

\begin{figure*}[]
   \centering
   \includegraphics[width=7.0in]{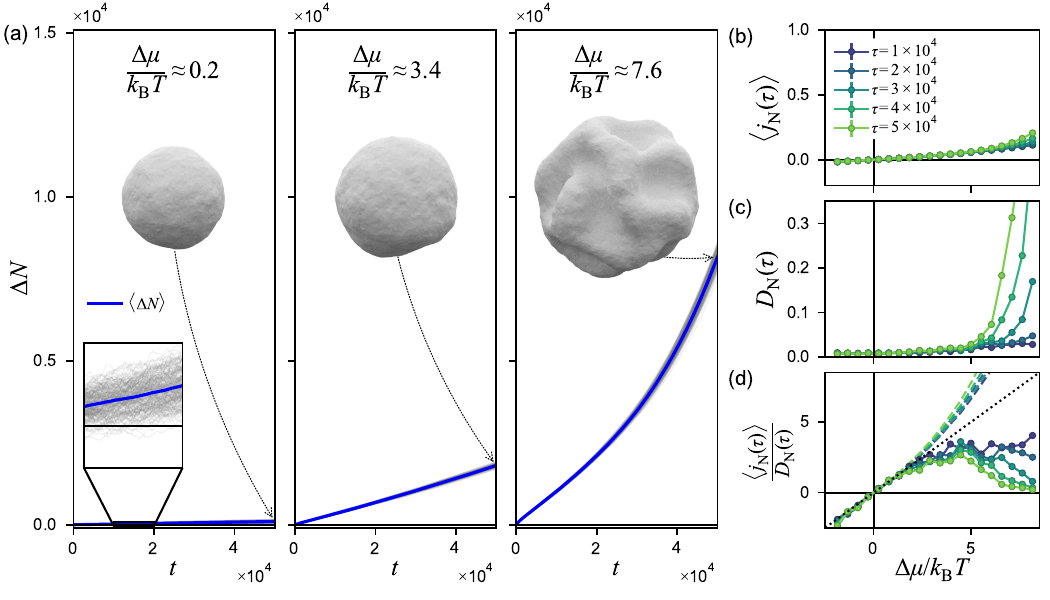}
   \caption{\textbf{Growth dynamics with slower insertion/removal attempt rate of $\bm{p_\mathrm{exchange}=0.01}$.}  (a) Ensembles of growth trajectories for varying values of excess chemical potential $\Delta\mu = \mu - \mu_\mathrm{eq}$, with the net influx of surface particles (vertices) $\Delta N = N(t) - N(0)$ plotted as a function of the number of elapsed Monte Carlo sweeps $t$. Light gray lines correspond to individual trajectories and blue lines correspond to the ensemble average $\langle\Delta N(t)\rangle$. (b-c) The average current $\langle j_\mathrm{N} (\tau)\rangle = \langle \Delta N(\tau)\rangle/\tau$, the diffusivity $D_\mathrm{N}(\tau) = \mathrm{Var}(\Delta N(\tau))/(2\tau)$, and the ratio $\langle j_\mathrm{N}(\tau)\rangle/D_\mathrm{N}(\tau)$ are shown as a function of the excess chemical potential $\Delta\mu$ for varying cutoff times $\tau$. In (d), the dashed lines show the same ratio if the denominator is replaced with the equilibrium diffusivity $D_\mathrm{N,eq} = D(\Delta\mu = 0)$, and the diagonal dotted line corresponds to the linear response prediction of Eq.\ 11 in the main text. For these simulations, the particle reservoir exchange attempt rate is $p_\mathrm{exchange} = 0.01$, the imposed osmotic pressure difference is $\Delta p = 0$, and the number of samples is $n_\mathrm{samples} = 200$. }
   \label{fig:fig_2_alt_dp0_pa01}
\end{figure*}

\begin{figure*}[]
   \centering
   \includegraphics[width=7.0in]{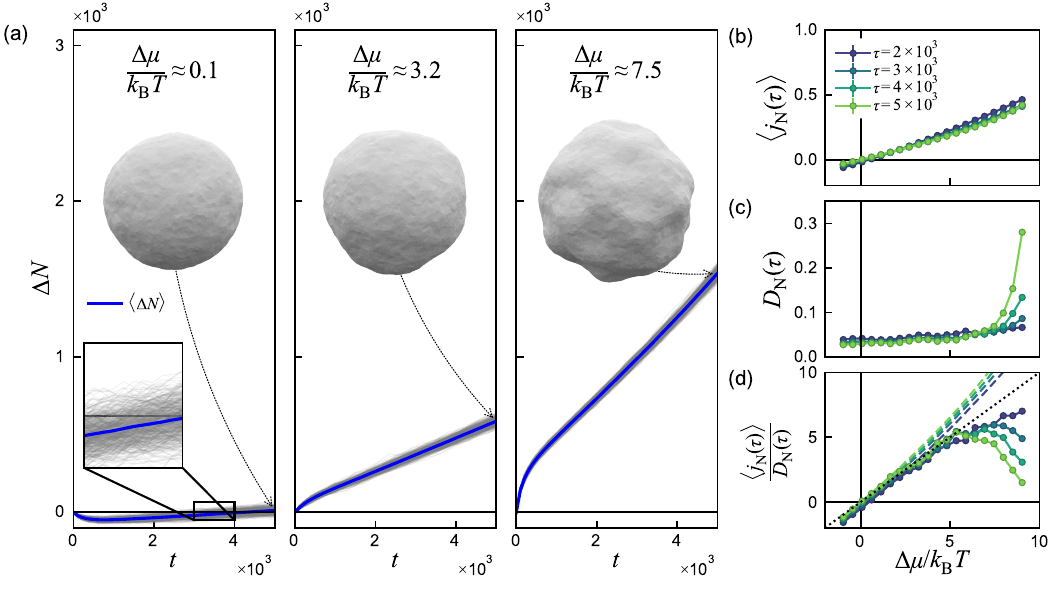}
   \caption{\textbf{Growth dynamics with a positive pressure difference of $\bm{\Delta p = 0.4}$.} (a) Ensembles of growth trajectories for varying values of excess chemical potential $\Delta\mu$, with the net influx of surface particles (vertices) $\Delta N = N(t) - N(0)$ plotted as a function of the number of elapsed Monte Carlo sweeps $t$. Here, the excess chemical potential is computed relative to the equilibrium chemical potential at finite pressure difference: $\Delta\mu = \mu - \mu_\mathrm{eq}\big|_{\Delta p}$. Light gray lines correspond to individual trajectories and blue lines correspond to the ensemble average $\langle\Delta N(t)\rangle$.  (b-c) The average current $\langle j_\mathrm{N} (\tau)\rangle = \langle \Delta N(\tau)\rangle/\tau$, the diffusivity $D_\mathrm{N}(\tau) = \mathrm{Var}(\Delta N(\tau))/(2\tau)$, and the ratio $\langle j_\mathrm{N}(\tau)\rangle/D_\mathrm{N}(\tau)$ are shown as a function of the excess chemical potential $\Delta\mu$ for varying cutoff times $\tau$. In (d), the dashed lines show the same ratio if the denominator is replaced with the equilibrium diffusivity $D_\mathrm{N,eq} = D(\Delta\mu = 0)$, and the diagonal dotted line corresponds to the linear response prediction of Eq.\ 11 in the main text. For these simulations, the particle reservoir exchange attempt rate is $p_\mathrm{exchange} = 1$, the imposed osmotic pressure difference is $\Delta p = 0.4$, and the number of samples is $n_\mathrm{samples} = 500$.}
   \label{fig:fig_2_alt_dp4_pa1}
\end{figure*}

\begin{figure*}[]
   \centering
   \includegraphics[width=7.0in]{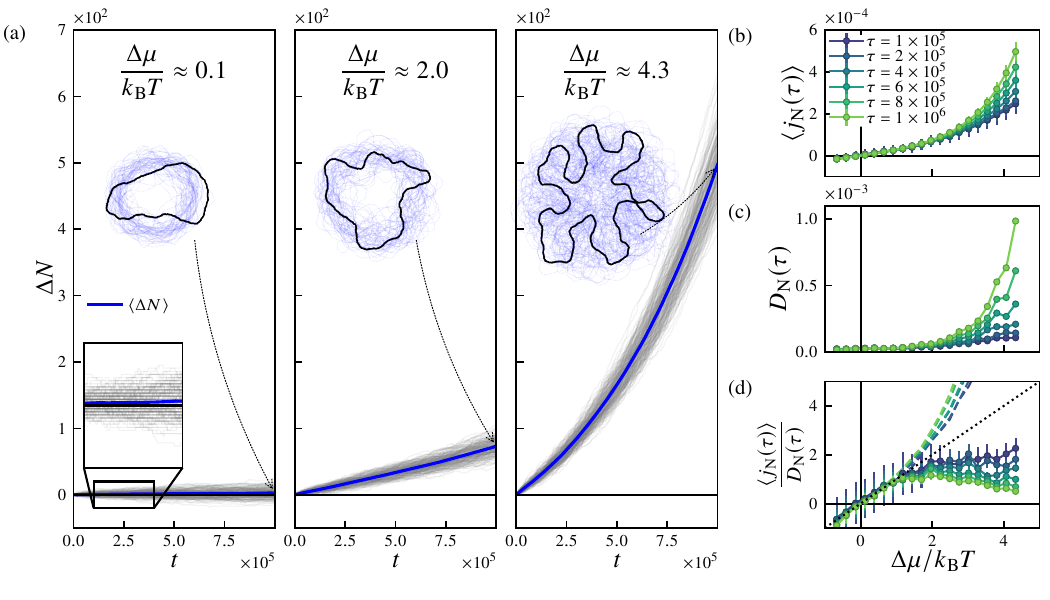}
   \caption{\textbf{Growth dynamics for two-dimensional systems.} (a) Ensembles of growth trajectories for varying values of excess chemical potential $\Delta\mu = \mu - \mu_\mathrm{eq}$, with the net influx of surface particles (vertices) $\Delta N = N(t) - N(0)$ plotted as a function of the number of elapsed Monte Carlo sweeps $t$.  Light gray lines correspond to individual trajectories and blue lines correspond to the ensemble average $\langle\Delta N(t)\rangle$. Representative vesicle configurations are shown in the insets, illustrating the transition from circular to highly deformed morphologies with increasing $\Delta\mu$.  (b-c) The average current $\langle j_\mathrm{N} (\tau)\rangle = \langle \Delta N(\tau)\rangle/\tau$, the diffusivity $D_\mathrm{N}(\tau) = \mathrm{Var}(\Delta N(\tau))/(2\tau)$, and the ratio $\langle j_\mathrm{N}(\tau)\rangle/D_\mathrm{N}(\tau)$ are shown as a function of the excess chemical potential $\Delta\mu$ for varying cutoff times $\tau$. In (d), the dashed lines show the same ratio if the denominator is replaced with the equilibrium diffusivity $D_\mathrm{N,eq} = D(\Delta\mu = 0)$, and the diagonal dotted line corresponds to the linear response prediction of Eq.\ 11 in the main text. For these simulations, the particle reservoir exchange attempt rate is $p_\mathrm{exchange} = 0.01$, the imposed osmotic pressure difference is $\Delta p = 0$, and the number of samples is $n_\mathrm{samples} = 200$. }
   \label{fig:fig_2_alt_2D_dp0_pa01}
\end{figure*}

\begin{figure*}[]
   \centering
   \includegraphics[width=7.0in]{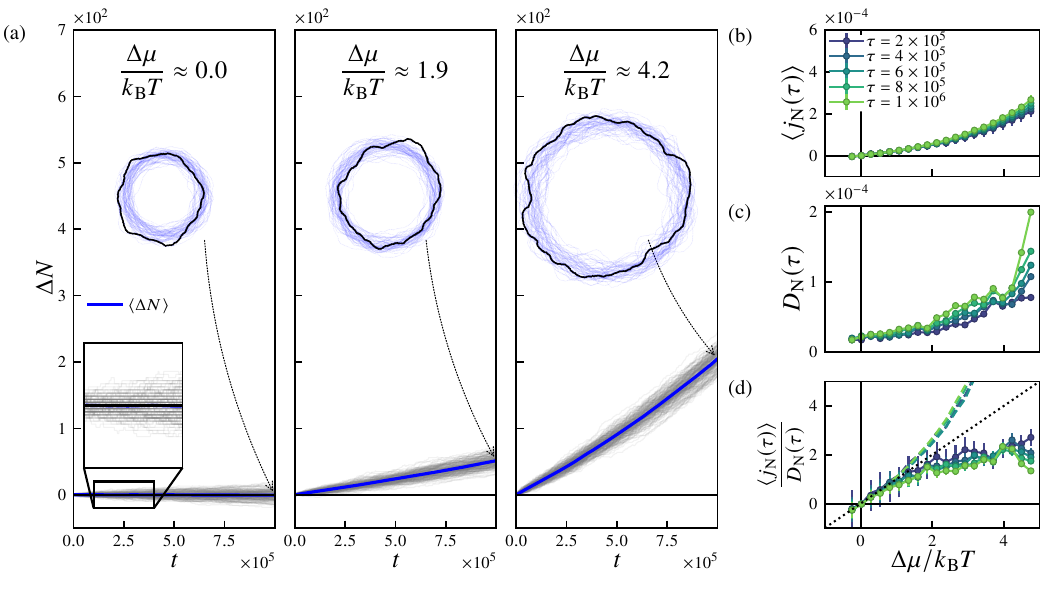}
   \caption{\textbf{Growth dynamics for two-dimensional systems with a positive pressure difference of $\bm{\Delta p = 0.05}$.} (a) Ensembles of growth trajectories for varying values of excess chemical potential $\Delta\mu$, with the net influx of surface particles (vertices) $\Delta N = N(t) - N(0)$ plotted as a function of the number of elapsed Monte Carlo sweeps $t$. Here, the excess chemical potential is computed relative to the equilibrium chemical potential at finite pressure difference: $\Delta\mu = \mu - \mu_\mathrm{eq}\big|_{\Delta p}$. Light gray lines correspond to individual trajectories and blue lines correspond to the ensemble average $\langle\Delta N(t)\rangle$. Representative vesicle configurations are shown in the insets, illustrating that vesicles maintain approximately circular morphologies even at large $\Delta\mu$ due to the stabilizing effect of internal pressure. (b-c) The average current $\langle j_\mathrm{N} (\tau)\rangle = \langle \Delta N(\tau)\rangle/\tau$, the diffusivity $D_\mathrm{N}(\tau) = \mathrm{Var}(\Delta N(\tau))/(2\tau)$, and the ratio $\langle j_\mathrm{N}(\tau)\rangle/D_\mathrm{N}(\tau)$ are shown as a function of the excess chemical potential $\Delta\mu$ for varying cutoff times $\tau$. In (d), the dashed lines show the same ratio if the denominator is replaced with the equilibrium diffusivity $D_\mathrm{N,eq} = D(\Delta\mu = 0)$, and the diagonal dotted line corresponds to the linear response prediction of Eq.\ 11 in the main text. For these simulations, the particle reservoir exchange attempt rate is $p_\mathrm{exchange} = 0.01$, the imposed osmotic pressure difference is $\Delta p = 0.05$, and the number of samples is $n_\mathrm{samples} = 200$. }
   \label{fig:fig_2_alt_2D_dp05_pa01}
\end{figure*}

\begin{figure*}[]
   \centering
   \includegraphics[width=7in]{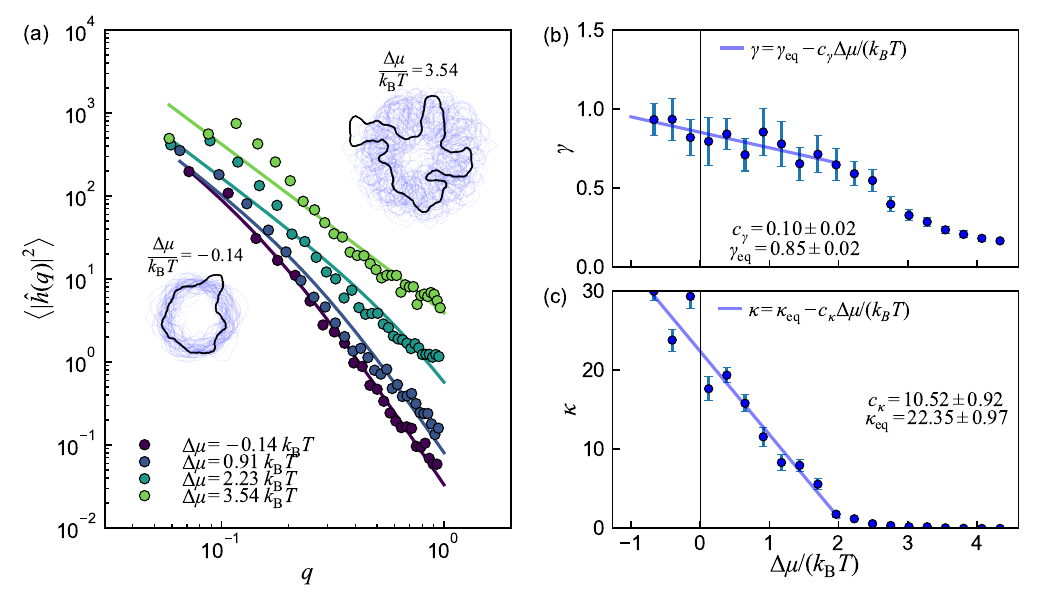}
   \caption{\textbf{Power spectra and renormalized mechanical properties of growing 2D vesicles at varying excess chemical potential.} (a) Mean squared amplitude $\langle |\hat{h}(q)|^2 \rangle$ of radial fluctuations of wave number $q$ for three different values of excess chemical potential $\Delta\mu$. Solid curves show fits to the theoretical model (Eq.\ A23 in the main text)  with $\Delta\mu$-dependent renormalized tension $\gamma$ and bending rigidity $\kappa$. Representative vesicle configurations are shown in the insets. (b-c) Variation of the renormalized parameters with $\Delta\mu/k_\mathrm{B}T$. The near-equilibrium regime ($\Delta\mu \approx 0$) shows approximately linear behavior characterized by slopes $c_\gamma$ and $c_\kappa$. Here, $p_\mathrm{exchange}=0.01$, $\Delta p = 0$, and $\tau=10^6$ sweeps.}
   \label{fig:fig_4_2D_dp0_pa01}
\end{figure*}

\begin{figure}[]
   \centering
   \includegraphics[width=3.375in]{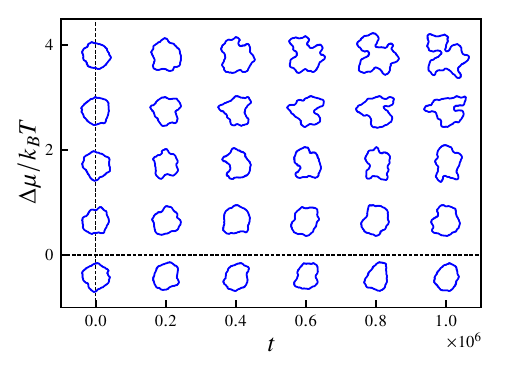}
   \caption{\textbf{Configurational dynamics of 2D vesicles for varied imposed excess chemical potential.} Representative snapshots of vesicle configurations after number of sweeps $t$, for different values of the excess chemical potential $\Delta\mu/k_\mathrm{B}T$ Images are centered on the corresponding values of $(t,\;\Delta\mu)$. For these simulations, the particle reservoir exchange attempt rate is $p_\mathrm{exchange} = 0.01$, the imposed osmotic pressure difference is $\Delta p = 0$, and the average initial number of particles is $\langle N_0 \rangle = 200$. }
   \label{fig:fig_3_2D_dp0_pa01}
\end{figure}

\begin{figure}[]
   \centering
   \includegraphics[width=3.375in]{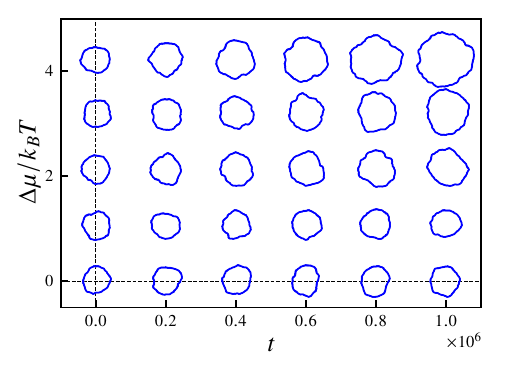}
   \caption{\textbf{Configurational dynamics of pressurized 2D vesicles for varied imposed excess chemical potential.} Representative snapshots of vesicle configurations after number of sweeps $t$, for different values of the excess chemical potential $\Delta\mu/k_\mathrm{B}T$ Images are centered on the corresponding values of $(t,\;\Delta\mu)$. For these simulations, the particle reservoir exchange attempt rate is $p_\mathrm{exchange} = 0.01$, the imposed osmotic pressure difference is $\Delta p = 0.05$, and the average initial number of particles is $\langle N_0 \rangle = 200$. }
   \label{fig:fig_3_2D_dp05pa01}
\end{figure}

\begin{figure}[]
   \centering
   \includegraphics[width=3.375in]{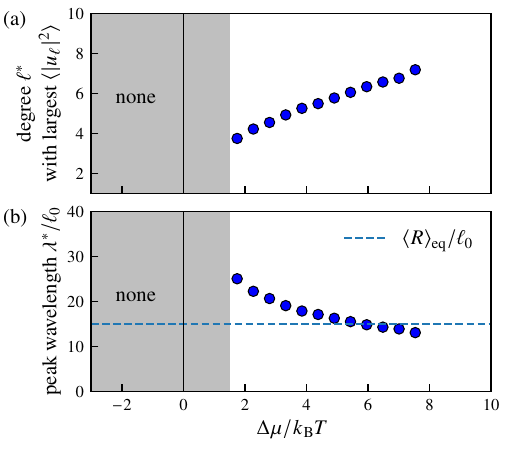}
    \caption{\textbf{Dependence of the characteristic spherical harmonic degree and wavelength on nonequilibrium driving.}  (a) Degree $\ell^*$ with largest mean squared amplitude $\langle|u_{\ell }|^2\rangle$ as a function of the excess chemical potential $\Delta\mu/k_\mathrm{B}T$. In the region shaded gray, no dominant mode exists. (b) Peak wavelength $\lambda^*$ vs. $\Delta\mu/k_\mathrm{B}T$, showing the characteristic length scale of shape fluctuations. }
   \label{fig:wavelength_vs_mu}
\end{figure}

\begin{figure}[]
  \centering
  \includegraphics[width=3.375in]{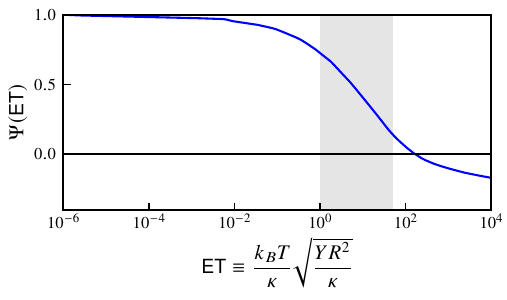}
\caption{\textbf{Renormalization group scaling function $\Psi(\mathsf{ET})$ for the buckling of externally pressurized elastic shells at finite temperature.} The scaling function $\Psi(\mathsf{ET})$, reproduced from Ref.\ 67 (main text), is plotted as a function of the elastothermal number $\mathsf{ET}$. The range of elastothermal numbers seen in our simulations is highlighted in gray.}\label{fig:psi} 
\end{figure}

\begin{figure}[]
   \centering
   \includegraphics[width=3.375in]{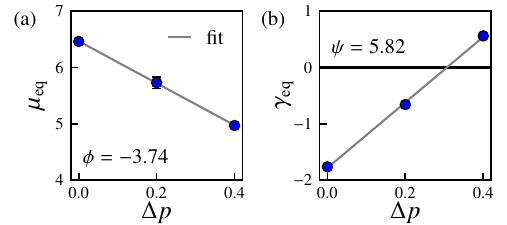}
   \caption{\textbf{Dependence of the equilibrium chemical potential and effective tension on the imposed osmotic pressure difference.} The (a) equilibrium chemical potential $\mu_\mathrm{eq}$ and (b) equilibrium tension $\gamma_\mathrm{eq}$ are plotted as functions of $\Delta p$. Data points with error bars show measured values, while solid lines correspond to fits to the expressions  $\mu_\mathrm{eq}\big|_{\Delta p} = \mu_\mathrm{eq}\big|_{\Delta p=0} + \phi \Delta p$ and $\gamma_\mathrm{eq}\big|_{\Delta p} = \gamma_\mathrm{eq}\big|_{\Delta p=0} + \psi \Delta p$, with slopes $\phi = -3.74$  (in units of $\ell_0^3$) and $\psi = 5.82$ (in units of $\ell_0$), respectively. For these data, $p_\mathrm{exchange}=1$, $\tau=5000$ sweeps, and the number of samples is $n_\mathrm{samples} = 500$.  }
   \label{fig:mu_eq_and_gamma_eq_vs_delta_p}
\end{figure}

\begin{figure}[]
   \centering
   \includegraphics[width=3.375in]{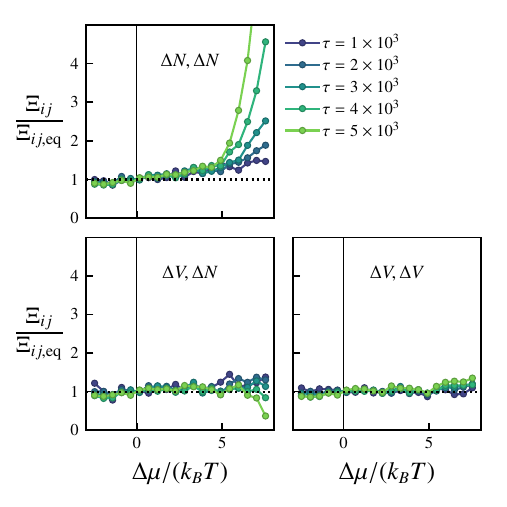}
   \caption{\textbf{How do the components of the flux covariance matrix  depend on cutoff time and distance to equilibrium?} The normalized component of $\Xi_{ij}/\Xi_{ij,\mathrm{eq}} = \mathrm{Cov}(J_i, J_j)/\mathrm{Cov}(J_i^\mathrm{eq}, J_j^\mathrm{eq})$ are plotted as functions of $\Delta\mu$ for varying cutoff times $\tau$. As $\bm{\Xi}$ is symmetric, only the lower triangular components are shown. For these data, $p_\mathrm{exchange}=1$, $\Delta p = 0$, and the number of samples is $n_\mathrm{samples} = 500$.}
   \label{fig:covariance_matrix_components}
\end{figure}